\documentclass{afu}
\usepackage[T1]{fontenc}
\usepackage{amsmath}
\usepackage{amssymb}
\usepackage{array}
\usepackage{tabularx}
\usepackage{makecell}
\usepackage{colortbl}
\usepackage{xurl}
\renewcommand{\email}[1]{\href{mailto:#1}{\nolinkurl{#1}}}
\renewcommand{\authorformat}[2][]{\mbox{\small\bfseries #2$^{#1}$}}
\renewcommand{\affiliationformat}[2][]{\mbox{\normalsize $^{#1}$#2}}
\usepackage{wrapfig}
\usepackage{needspace}
\newcommand{\runningtitle}{SkinMind: Human-Inspired Graph Probing for Dermatology AI}
\AddToHook{cmd/maketitle/after}{\fancyhead[L]{\nouppercase{\small\textbf{\runningtitle}}}}
\setcitestyle{numbers,square}
\providecommand{\Description}[1]{}

\definecolor{skinmint}{HTML}{DCEDE8}
\definecolor{skinteal}{HTML}{4F8F83}
\definecolor{skinrose}{HTML}{F5E7E5}
\definecolor{skincoral}{HTML}{C97878}
\definecolor{referencebg}{HTML}{F5F5F5}
\definecolor{referenceband}{HTML}{E6E6E6}
\definecolor{gptpolicybg}{HTML}{F0F7F4}
\definecolor{gptpolicyband}{HTML}{DCEDE8}
\definecolor{geminipolicybg}{HTML}{F1F3FA}
\definecolor{geminipolicyband}{HTML}{E2E7F4}
\definecolor{sectionbg}{RGB}{231,240,250}
\definecolor{sectiontext}{RGB}{31,78,121}
\definecolor{baselinebg}{RGB}{245,245,245}
\definecolor{oursbg}{RGB}{235,246,255}
\definecolor{deltabg}{RGB}{255,242,242}
\definecolor{deltared}{RGB}{200,35,35}

\usepackage{listings}

\newcolumntype{Y}{>{\raggedright\arraybackslash}X}
\newcolumntype{R}{>{\raggedleft\arraybackslash}X}
\title{SkinMind: Equipping Agentic RAG with Human-Inspired Graph Probing for Reliable Dermatology AI}
\date{November 6, 2026}
\author[1,2]{Jing Hao}
\author[1]{Weinan Zhou}
\author[1]{Zhihong Lu}
\author[1,3]{Yaling Shen}
\author[1,4]{Yiming Xu}
\author[1]{Yefeng Liu}
\author[3]{Siyuan Yan}
\author[6]{Shuaibo Li}
\author[1,5]{Ming Dai}
\author[1]{Junjun He}
\author[1]{Zuyi Zhu}
\author[1]{Jinjie Gu}
\author[1,*]{Lijun Wu}
\author[1,*]{Jian Wang}

\affiliation[1]{Ant Group, AFu Team}
\affiliation[2]{The University of Hong Kong}
\affiliation[3]{Monash University}
\affiliation[4]{University of Science and Technology of China}
\affiliation[5]{Southeast University}
\affiliation[6]{The Hong Kong University of Science and Technology (Guangzhou)}
\contribution[*]{Corresponding authors.}
\correspondence{Lijun Wu, \email{lijun_wu@outlook.com}; Jian Wang, \email{bobblair.wj@antgroup.com}}

\abstract{
Reliable dermatology AI requires access to verifiable evidence distributed across heterogeneous textual and visual sources. However, conventional retrieval-augmented generation (RAG) typically relies on one-shot retrieval from flat corpora, limiting its ability to adapt retrieval granularity, explore multi-hop relations, and integrate multimodal clinical evidence. We introduce \textbf{SkinMind}, a multimodal agentic RAG framework that equips policy models with human-inspired graph probing for autonomous evidence seeking. SkinMind organizes retrieval as an iterative interaction between an orchestrator and a graph environment. Through eight hierarchical graph-probing tools, the orchestrator progressively identifies relevant entities, inspects node-level information, traverses relations and paths, and retrieves source-text and visual evidence. This progressive-disclosure design allows the model to adaptively determine the retrieval modality, exploration direction, and evidence granularity while constructing a query-specific evidence subgraph. To support autonomous graph probing, we construct \textbf{SkinMind-KG} from a 62M-token corpus of dermatology guidelines and textbooks and 423K images, guided by a dermatologist-informed schema and a conflict-resolution mechanism. The resulting graph contains over 1.0M nodes and 2.0M edges, unifying textual, relational, and visual evidence for retrieval. Across six dermatology VQA benchmarks and two policy-model settings, SkinMind improves the average score over direct answering by \(8.70\) and \(8.89\) points with GPT-5.4-mini and Gemini-2.5-Flash, respectively, while consistently outperforming competing RAG methods. Together, SkinMind and SkinMind-KG provide an effective foundation for evidence-grounded multimodal reasoning in dermatology.
}
\metadata[Keywords]{Retrieval-augmented generation, agentic search, knowledge graphs, multimodal reasoning, dermatology}
\hypersetup{pdftitle={SkinMind: Equipping Agentic RAG with Human-Inspired Graph Probing for Reliable Dermatology AI},pdfauthor={Jing Hao, Weinan Zhou, Zhihong Lu, Yaling Shen, Yiming Xu, Yefeng Liu, Siyuan Yan, Shuaibo Li, Ming Dai, Junjun He, Zuyi Zhu, Jinjie Gu, Lijun Wu, Jian Wang}}
\begin{document}
\maketitle

\section{Introduction}

Multimodal large language models (MLLMs) have demonstrated remarkable capabilities in language understanding, visual perception, and multi-step reasoning, showing considerable potential for addressing complex real-world tasks~\cite{openai2026gpt55,comanici2025gemini25,googledeepmind2026gemini31pro,openai2026gpt54mini,team2025kimi,xiaomi2025mimo,abdin2024phi}. Dermatology represents a particularly promising application: many skin conditions are visually observable, and patients increasingly seek convenient access to professional medical guidance through online consultations. Accordingly, medical and dermatology-specific MLLMs have shown potential across tasks such as visual question answering, diagnostic decision-making, and treatment recommendation~\cite{fan2026oralgpt,hao2026oralgpt,hao2026towards,chen2026skingpt,shen2025skingpt,ru2026dermogpt,zhou2024skingpt4,liu2026dermagent}. Despite this promise, their deployment in safety-critical healthcare settings remains challenging. First, they may generate fluent but factually incorrect statements, potentially leading to harmful clinical decisions. Second, their parametric knowledge is acquired from static training corpora and cannot be readily updated as clinical evidence and guidelines evolve. Third, their responses often lack verifiable references to supporting medical evidence, substantially limiting their reliability and interpretability.

\begin{wrapfigure}{R}{0.49\textwidth}
    \centering
    \captionsetup{font=small,skip=4pt}
    \includegraphics[width=\linewidth]{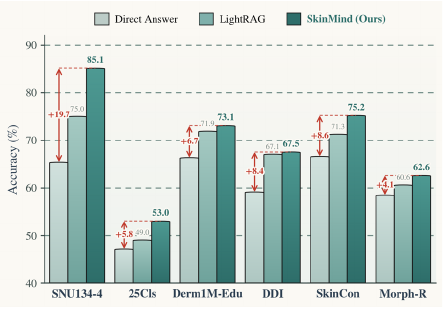}
    \caption{Performance comparison across six dermatology VQA benchmarks using Gemini-2.5-Flash as the policy model. SkinMind consistently outperforms direct answering and LightRAG.}
    \label{fig:overview}
    \Description{Bar charts comparing SkinMind, direct answering, and LightRAG across six dermatology visual question answering benchmarks using Gemini-2.5-Flash.}
\end{wrapfigure}

Retrieval-augmented generation (RAG) offers a promising means of mitigating these limitations by grounding model responses in external, authoritative, and traceable medical evidence~\cite{graphrag,du2026arag,guo2024lightrag,csakar2025maximizingrag,wang2025mirarag,lu2025doctorrag,wu2025medicalrag,madavan2025med,ma2025tog2,gutierrez2024hipporag,asai2024selfrag,hao2026oralagent}. Conventional RAG systems, however, typically perform a one-shot retrieval over a flat collection of textual chunks, presenting two limitations for complex medical reasoning~\cite{han2025rag}. First, one-shot retrieval relies solely on the initial query and cannot adapt to information needs that emerge during multi-step reasoning. Second, flat-chunk retrieval often provides fragmented evidence with insufficient relational context while introducing irrelevant information. Addressing the first limitation requires adaptive, multi-round retrieval, whereas addressing the second calls for a structured representation of external medical knowledge.

Recent agentic RAG methods seek to address the first limitation by interleaving reasoning with iterative retrieval~\cite{du2026arag,besrour2025ragenta}. Rather than relying on a fixed set of documents retrieved upfront, the model dynamically determines when additional evidence is needed and formulates new queries based on its intermediate reasoning state. Nevertheless, because these methods typically operate over isolated text chunks, they do not fully resolve the fragmentation and loss of relational context inherent in flat-text retrieval.
Graph-based RAG improves upon flat-text retrieval by organizing knowledge into entities and relations~\cite{graphrag,guo2024lightrag,zhao2025medrag}. However, many existing methods still treat the graph primarily as a retrieval index, selecting entities or subgraphs through embedding similarity or predefined expansion rules. Although the knowledge is graph-structured, the graph itself is rarely exposed to the model for interactive exploration. This contrasts with a human-inspired evidence-seeking process, in which a reasoner begins with a relevant concept, inspects its relations, follows informative paths, and adapts the exploration as new information emerges. This process follows the principle of \emph{progressive disclosure}~\cite{springer2020progressive}, whereby lightweight structural cues guide navigation, while detailed descriptions or source content are accessed only when needed. Compared with retrieving a fixed candidate set, such navigation allows the retrieval path, scope, and granularity to evolve with the reasoning process, facilitating the flexible construction of query-specific evidence subgraphs while reducing irrelevant context.

Inspired by this human-like evidence-seeking process, we propose \textbf{SkinMind}, a multimodal agentic RAG framework for autonomous and reliable dermatology assistance. Rather than treating the knowledge graph as a static retrieval index, SkinMind exposes it as an interactive environment and equips the agent with eight specialized graph-probing tools for interleaved reasoning and retrieval of textual, relational, and visual evidence. Given a clinical query, the agent autonomously identifies relevant entry nodes, inspects their relations, follows informative paths, and progressively accesses node descriptions, source content, and clinical images as needed. To provide a comprehensive and traceable foundation of domain-specific evidence, we construct \textbf{SkinMind-KG}, a large-scale heterogeneous dermatology knowledge graph containing over \(1\) million nodes and \(2\) million edges across 23 core entity types and 25 core relation types. 
Notably, SkinMind-KG explicitly links \(301{,}561\) images to disease entities, providing rich visual–conceptual associations for multimodal evidence seeking.

We conduct extensive experiments on six public dermatology VQA benchmarks spanning morphology recognition, disease diagnosis, and clinical reasoning in both closed- and open-ended settings. As illustrated in Fig.~\ref{fig:overview}, SkinMind consistently outperforms direct answering across all six benchmarks, with gains of \(4.1\)–\(19.7\) points under Gemini-2.5-Flash. More broadly, across both policy-model settings, it surpasses LightRAG by an average of \(5.75\) points with GPT-5.4-mini and \(3.60\) points with Gemini-2.5-Flash. These results suggest that adaptive graph exploration can support more effective use of external knowledge and improve multimodal clinical reasoning. Our main contributions are summarized as follows:
\begin{itemize}
    \item We propose \textbf{SkinMind}, a multimodal agentic RAG framework that treats a knowledge graph as an interactive environment, enabling interleaved reasoning and adaptive evidence seeking through eight graph-probing tools.
    \item We construct \textbf{SkinMind-KG}, a large-scale heterogeneous dermatology knowledge graph comprising over \(1\) million nodes and \(2\) million edges, with more than \(300{,}000\) images explicitly linked to disease entities.
    \item Extensive experiments on six dermatology VQA benchmarks across two policy-model settings demonstrate that SkinMind consistently outperforms direct answering and competing RAG methods, validating the effectiveness of adaptive graph exploration for evidence-grounded multimodal clinical reasoning.
\end{itemize}

\section{Related Work}
\label{sec:related_work}

\paragraph{Retrieval-Augmented Generation.}
RAG augments language models with evidence from external
corpora~\cite{lewis2020rag}. Adaptive methods interleave reasoning and
retrieval, respond to generation uncertainty, incorporate self-critique,
or route queries among retrieval
strategies~\cite{trivedi2023ircot,jiang2023flare,asai2024selfrag,jeong2024adaptiverag}.
GraphRAG, LightRAG, and HippoRAG organize evidence through entity graphs,
community summaries, and associative
retrieval~\cite{graphrag,guo2024lightrag,gutierrez2024hipporag}.
Think-on-Graph 2.0 alternates graph and document retrieval, while Medical
Graph RAG integrates medical sources and controlled
vocabularies~\cite{ma2025tog2,wu2025medicalrag}. Agentic approaches give
models control over evidence acquisition: A-RAG exposes hierarchical
passage-search interfaces, whereas RAGentA coordinates retrieval, answer
generation, and verification across
agents~\cite{du2026arag,besrour2025ragenta}. SkinMind combines these
directions by exposing a heterogeneous multimodal graph as an interactive
environment. Its policy model selectively accesses nodes, relations,
paths, source passages, and images at the granularity needed for each
question.

\paragraph{Multimodal Models in Dermatology.}
SkinGPT-4 adapts a visual language model for interactive dermatology,
while PanDerm provides domain-specific visual
pretraining~\cite{zhou2024skingpt4,yan2025panderm}. Derm1M and MM-Skin
supply large image--text resources for cross-modal representation,
retrieval, and dermatology-specific vision--language
modeling~\cite{yan2025derm1m,zeng2025mmskin}. DermoGPT emphasizes
morphology-grounded diagnostic reasoning~\cite{ru2026dermogpt}.
Tool-augmented systems complement these models with external memory and
specialized components. SkinGPT-X coordinates specialist agents with an
evolving case memory, while DermAgent combines perception tools, case
and guideline retrieval, disease taxonomy, and evidence
auditing~\cite{chen2026skingpt,liu2026dermagent}. SkinMind uses a
source-linked dermatology graph as its central retrieval environment,
allowing the policy model to select the modality, relation, and evidence
granularity during inference. This external knowledge can be updated
independently of the policy model and traced to its supporting sources.

\section{SkinMind}

\begin{figure}[!htbp]
    \centering
    \includegraphics[width=\textwidth]{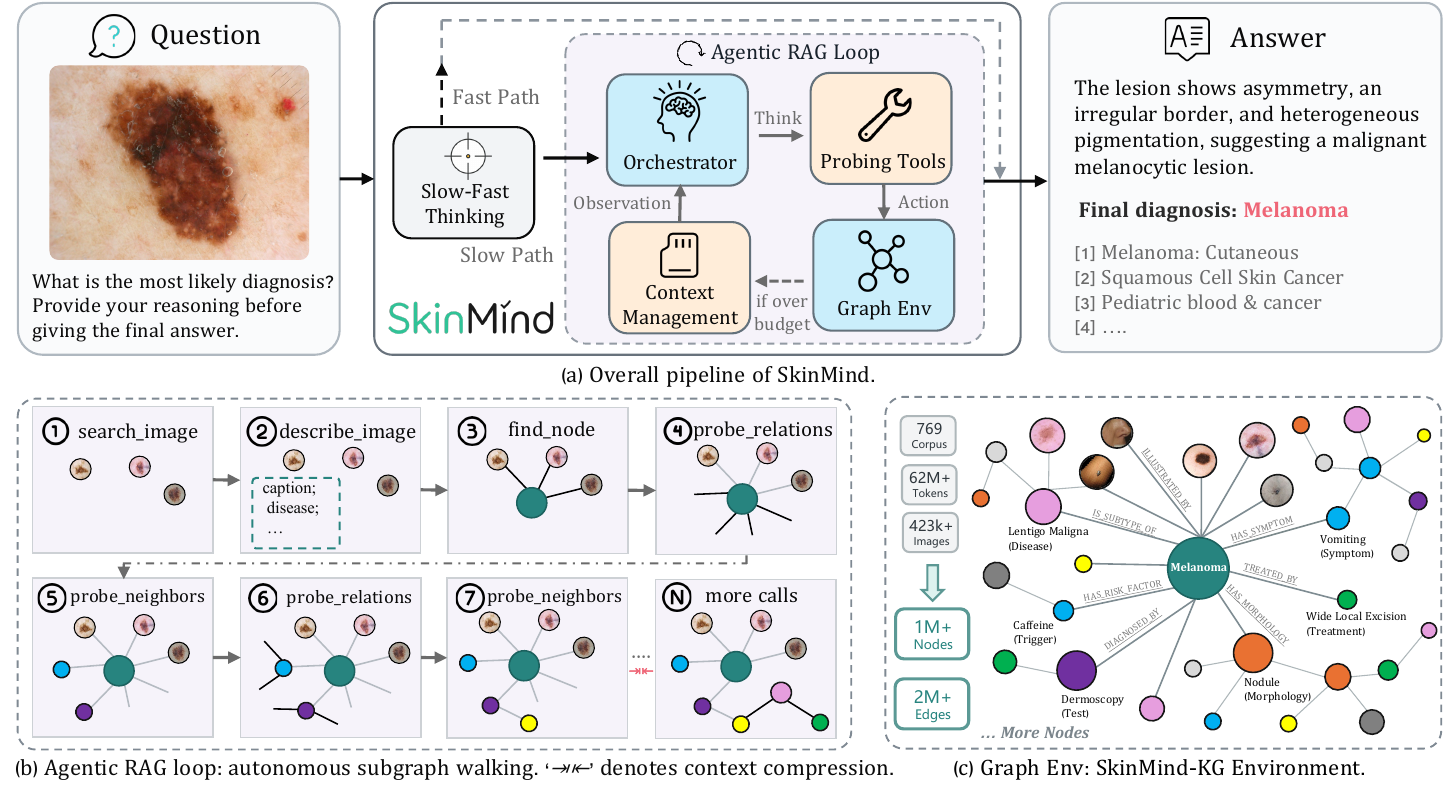}
    \caption{Overview of SkinMind. The orchestrator adaptively explores SkinMind-KG through hierarchical graph-probing tools to construct query-specific evidence subgraphs for response generation.}
    \label{fig:framework}
    \Description{SkinMind architecture and graph-probing interfaces, showing the orchestrator, slow and fast reasoning paths, graph environment, and iterative retrieval of textual and visual evidence.}
\end{figure}

\subsection{Overview}
\label{sec:overview}
SkinMind is a multimodal agentic RAG framework equipped with a specialized suite of graph-probing tools for interleaved reasoning and evidence retrieval over a dermatology-specific graph environment. Given a clinical query, optionally accompanied by an image, SkinMind first routes the query through a slow--fast reasoning mechanism. Queries that can be answered reliably using the orchestrator's internal knowledge are handled by the fast path, whereas queries requiring external evidence, multimodal retrieval, or multi-step reasoning are directed to the slow path. The slow path implements an iterative Agentic RAG loop with four components: an \emph{orchestrator}, a suite of \emph{graph-probing tools}, a graph-based \emph{environment}, and a \emph{context manager}. The orchestrator acts as the central policy model: it reasons over the current context, selects and invokes appropriate tools, and incorporates the returned observations into subsequent decisions. Once sufficient evidence has been collected, it terminates the interaction loop and generates an evidence-grounded response for the user.

\subsection{Slow--Fast Thinking}
Let \(Q\) denote a textual clinical query and \(I\) an optional clinical image. A routing policy first determines whether external evidence seeking is necessary:
\begin{equation}
    r = \pi_{\mathrm{route}}(Q,I),
    \qquad
    r \in \{\textsc{Fast},\textsc{Slow}\}.
    \label{eq:routing}
\end{equation}
The routing decision accounts for the user intent, query complexity, and the need for external or multimodal evidence. If \(r=\textsc{Fast}\), the orchestrator directly generates a response. Otherwise, the query enters the graph-probing loop described below.

\subsection{Agentic Graph-Probing Loop}
\paragraph{Loop Formulation.}

For a query assigned to the slow path, the orchestrator alternates between
reasoning, tool invocation, and answer generation based on the
accumulated interaction history. At step \(t\), the orchestrator state
is defined as
\begin{equation}
    s_t = \left(Q,I,\mathcal{H}_t\right),
    \label{eq:agent_state}
\end{equation}
where \(Q\) is the textual query, \(I\) is an optional clinical image,
and
\begin{equation}
    \mathcal{H}_t =
    \bigl((a_0,o_0),\ldots,(a_{t-1},o_{t-1})\bigr)
    \label{eq:interaction_history}
\end{equation}
denotes the interaction history before step \(t\). Here, \(a_i\) is the
action taken at step \(i\), and \(o_i\) is the corresponding observation
returned by the graph environment.

Conditioned on \(s_t\), the orchestrator either invokes a graph-probing
tool or terminates the loop. For a non-terminal action, the selected
tool operates on the graph environment and returns an observation,
which is appended to the interaction history for the next reasoning
step. The loop continues until the orchestrator selects
\textsc{Finish} or reaches the maximum interaction budget
\(L_{\max}\), producing a multi-turn trajectory
\begin{equation}
\begin{split}
    \tau = \bigl(
    &(s_0,a_0,o_0),
    (s_1,a_1,o_1),
    \ldots, \\
    &(s_{L-1},a_{L-1},o_{L-1}),
    (s_L,\textsc{Finish},Y)
    \bigr),
\end{split}
\label{eq:trajectory}
\end{equation}
where \(L\leq L_{\max}\) and \(Y\) denotes the final response.

\paragraph{Tool Calling.}
Let \(\mathcal{T}=\{T_1,\ldots,T_8\}\) denote the suite of
graph-probing tools operating on textual and visual information in the
graph environment. At each step, the orchestrator selects either a
parameterized tool call
\begin{equation}
    a_t=(T_{k_t},\phi_t), \qquad T_{k_t}\in\mathcal{T},
\end{equation}
where \(\phi_t\) denotes the tool arguments, or the terminal action
\(\textsc{Finish}\). This decision is conditioned on the current state:
\begin{equation}
    a_t \sim \pi_{\theta}\!\left(\cdot\mid s_t\right).
    \label{eq:policy}
\end{equation}
The definitions and functionalities of these tools are detailed under
\emph{Hierarchical Graph-Probing Interfaces}.

\paragraph{Environment.}
Let \(\mathcal{E}\) denote \textbf{SkinMind-KG}, our
dermatology-specific multimodal graph environment. For each
non-terminal action, the environment executes the corresponding tool
call and returns an observation:
\begin{equation}
    o_t = \mathcal{E}(a_t),
    \qquad a_t \neq \textsc{Finish}.
    \label{eq:environment}
\end{equation}
The construction and organization of SkinMind-KG are detailed under
\emph{SkinMind-KG Construction and Statistics}.

\paragraph{Context Management.}
To control context growth during multi-step graph exploration, SkinMind
monitors both individual tool observations and the accumulated context.
If a single observation exceeds 4{,}096 tokens, the orchestrator compresses
it before incorporating it into the interaction history. If the
accumulated context exceeds 16{,}384 tokens, the orchestrator summarizes
earlier interactions to reduce the overall context length. These
operations preserve query-relevant evidence and retrieval provenance
required for subsequent reasoning and graph exploration.

\subsection{Hierarchical Graph-Probing Interfaces}
\label{sec:tool_interfaces}
SkinMind equips the orchestrator with eight graph-probing tools for
retrieving and exploring textual, relational, and visual evidence in
SkinMind-KG. As summarized in Table~\ref{tab:retrieval_tools}, six tools
support textual retrieval and graph exploration, while two tools support
visual retrieval and inspection. Following the principle of
\emph{progressive disclosure}~\cite{springer2020progressive,du2026arag},
the interfaces initially return compact
retrieval results and expose more detailed attributes, relations, and
source evidence only when requested, thereby reducing unnecessary
context overhead.

\begin{table}[!htbp]
\caption{\textbf{Overview of SkinMind's graph-probing tool suite.} The first six tools support textual retrieval and relation-aware graph exploration, while the last two support visual retrieval and metadata inspection.}
\label{tab:retrieval_tools}
\centering
\footnotesize
\setlength{\tabcolsep}{5pt}
\renewcommand{\arraystretch}{1.12}
\begin{tabularx}{\linewidth}{@{}l >{\raggedright\arraybackslash}X >{\raggedright\arraybackslash}X >{\raggedright\arraybackslash}X@{}}
\toprule
\textbf{Tool} & \textbf{Function} & \textbf{Arguments} & \textbf{Output} \\
\midrule
\texttt{find\_node}
& Retrieve graph nodes using a textual query
& Query, top-$K$
& Node IDs, names, and types \\

\texttt{read\_node}
& Retrieve the structured information of a node
& Node ID
& Node attributes and description \\

\texttt{read\_chunks}
& Retrieve source-text chunks associated with a node
& Node ID, chunk limit, character limit
& Source-text chunks \\

\texttt{probe\_relations}
& List relation types associated with a node
& Node ID
& Relation types and neighbor counts \\

\texttt{probe\_neighbors}
& Expand a node along a specified relation
& Node ID, relation, top-$K$
& Neighbor IDs, names, and types \\

\texttt{probe\_path}
& Search for relational paths between two nodes
& Source ID, target ID, hop limit
& Node and relation sequences \\

\texttt{search\_image}
& Retrieve image nodes using a text or image query
& Text/image query, top-$K$
& Image-node IDs and metadata \\

\texttt{describe\_image}
& Retrieve the metadata of an image node
& Image-node ID
& Caption, morphology, and disease labels \\
\bottomrule
\end{tabularx}

\end{table}

The orchestrator can enter the graph through textual node retrieval with
\texttt{find\_node} or through text-to-image or image-to-image retrieval
with \texttt{search\_image}. For image-to-image retrieval,
\texttt{search\_image} encodes the query image with
DermLIP~\cite{yan2025derm1m} and searches the image-node index by cosine
similarity, returning image identifiers, similarity scores, and source
types. Depending on its information needs, the orchestrator may inspect
captions, clinical attributes, and aligned disease labels using
\texttt{describe\_image}. The image-to-disease associations established
during graph construction allow these labels to serve as anchors for
locating \texttt{Disease} nodes through \texttt{find\_node}.

To gather further evidence, the orchestrator can inspect node details
using \texttt{read\_node}, explore graph structure using
\texttt{probe\_relations}, \texttt{probe\_neighbors}, and
\texttt{probe\_path}, and access fine-grained source evidence using
\texttt{read\_chunks}. Through iterative tool calls, it adaptively
determines the retrieval modality, exploration direction, and level of
evidence granularity based on its evolving information needs. The
retrieved nodes, relations, source chunks, and image records collectively
form a query-specific evidence subgraph for subsequent reasoning and
response generation. Figure~\ref{fig:framework}(b) illustrates an example
graph-probing trajectory.

\section{SkinMind-KG Construction and Statistics}
\label{sec:graph_construction}

\begin{figure}[!htbp]
    \centering
    \includegraphics[width=\textwidth]{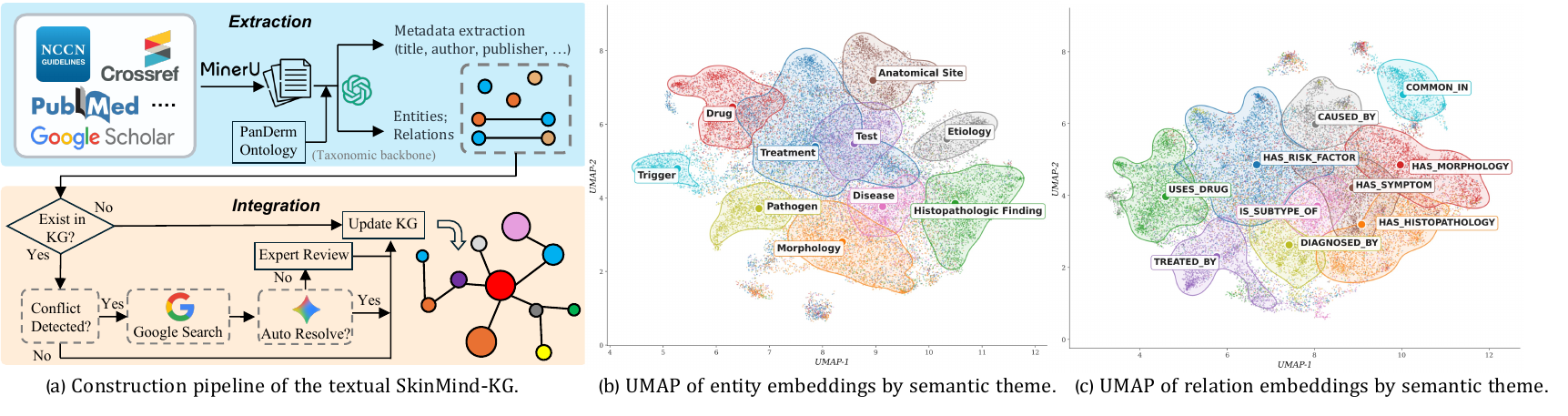}
    \caption{Construction and semantic organization of SkinMind-KG. The ontology-guided pipeline integrates knowledge with conflict resolution, while UMAP projections of entity and relation embeddings reveal clusters across semantic categories.}
    \label{fig:kg_construction}
    \Description{Construction of SkinMind-KG from dermatology texts and images, together with the organization and statistics of the resulting multimodal knowledge graph.}
\end{figure}

\subsection{Knowledge Graph Construction}
We construct the textual component of SkinMind-KG from a large-scale
dermatology corpus comprising 227 clinical guidelines and 542 textbooks
published since 2015. We collected dermatology guidelines and textbooks
from international online sources, selecting the latest available
guideline versions and textbook editions at the time of collection.
We parse the raw documents using
MinerU~\cite{niu2026mineru2} and employ a lightweight LLM to extract
document-level metadata, including titles, authors, publishers,
publication years, and editions. To establish an authoritative disease taxonomy, we adopt the PanDerm
ontology~\cite{yan2025panderm}, which comprises 127 disease concepts and their
curated hierarchical relationships. We encode these relationships as
\texttt{IS\_SUBTYPE\_OF} relations and preload the resulting hierarchy
into SkinMind-KG as its initial taxonomic backbone. Through iterative
consultations with senior dermatologists, we further define a schema
containing 23 entity types and 25 relation types to guide information
extraction.
Following LightRAG~\cite{guo2024lightrag}, we segment the approximately
62M token corpus into chunks with a length of 1,200
tokens. GPT-5.4-mini is then used to extract entities, relations, and
their descriptions from each chunk according to the predefined schema.
During corpus extraction, the initial disease hierarchy can be extended
with additional disease entities and subtype relations.

Entity and relation extraction is restricted to statements explicitly
supported by the source text. The source context guides the interpretation
of clinical qualifiers and the polarity of these statements. Extracted
entities and textual relations retain links to their supporting source
chunks, which preserve document-level provenance. Through these links,
the orchestrator can retrieve supporting passages and their document
identifiers using \texttt{read\_chunks} when node-level descriptions are
insufficient, providing access to the original clinical context.

\paragraph{Conflict Detection and Resolution.}
Integrating heterogeneous sources may introduce contradictory claims
about the same entity or relation. We therefore incorporate a
\emph{detect-then-resolve} procedure into the construction pipeline. A
conflict is identified only when two claims make contradictory
assertions about the same fact under the same clinical context. Claims
associated with different clinical contexts are retained as
context-specific evidence rather than treated as conflicts. For each detected conflict, Gemini-3.1-Pro~\cite{googledeepmind2026gemini31pro} formulates targeted queries
and retrieves relevant evidence through Google Search. It then evaluates
the retrieved evidence to determine whether the conflict can be
reliably resolved. Resolvable cases are automatically reconciled,
whereas ambiguous cases are referred to domain experts. In total, the
procedure processed 1,375 description-level conflicts, automatically
resolving 1,366 cases (99.3\%) and referring the remaining nine for
expert review. The overall procedure is illustrated in
Figure~\ref{fig:kg_construction}(a).

\begin{wraptable}{R}{0.51\textwidth}
\captionsetup{font=small,skip=4pt}
\caption{Node and edge counts, degree distribution, and connected-component
statistics of SkinMind-KG after multimodal integration.}
\label{tab:kg_stats}
\centering
\footnotesize
\begin{tabularx}{\linewidth}{@{}>{\raggedright\arraybackslash}Xr@{}}
\toprule
\textbf{Property} & \textbf{Value} \\
\midrule
\multicolumn{2}{@{}l}{\textit{Scale \& Composition}} \\
Nodes & 1{,}006{,}079 \\
Edges & 2{,}038{,}419 \\
Graph type & Undirected, unweighted \\
99\%-coverage entity types & 23 \\
99\%-coverage relation types & 25 \\
Image nodes & 423{,}016 (42.0\%) \\
\quad Linked to disease nodes & 301{,}561 (71.3\%) \\
\quad Isolated image nodes & 121{,}455 (28.7\%) \\
\midrule
\multicolumn{2}{@{}l}{\textit{Degree \& Density}} \\
Average degree & 4.05 \\
Median / p90 / p99 degree & 1 / 5 / 47 \\
Non-isolated nodes & 819{,}864 (81.5\%) \\
Maximum degree & 64{,}345 \\
Degree-one nodes & 432{,}054 (52.7\%) \\
Graph density & $4.0\times10^{-6}$ \\
\midrule
\multicolumn{2}{@{}l}{\textit{Connectivity}} \\
Connected components & 191{,}613 \\
Isolated nodes & 186{,}215 (18.5\%) \\
LCC (nodes) & 805{,}181 (80.0\%) \\
LCC (edges) & 2{,}029{,}050 (99.5\%) \\
\bottomrule
\end{tabularx}
\par\smallskip
\raggedright
\emph{Note:} LCC denotes the largest connected component. Degree
percentiles and the proportion of degree-one nodes use non-isolated
nodes. Linked and isolated image percentages use all image nodes.
Core type coverage excludes image nodes and \texttt{ILLUSTRATED\_BY}
edges, respectively.
\end{wraptable}

\paragraph{Visual Modality Integration.}
After constructing the textual graph, we augment SkinMind-KG with
423,016 image nodes from Derm1M~\cite{yan2025derm1m}, most of which are
accompanied by disease labels. To integrate these images into the graph,
we align their disease labels with textual graph nodes of type
\texttt{Disease} through a three-stage procedure. First, normalized
image labels are matched exactly against the names of disease nodes.
Second, for unmatched labels, Qwen3-Embedding-4B~\cite{zhang2025qwen3} is
used to retrieve the top-10 semantically closest disease nodes. Third,
GPT-5.4-mini disambiguates among these candidates and determines whether
a reliable label-to-node alignment can be established. Based on the
resulting alignments, image nodes are connected to the corresponding
disease nodes through \texttt{ILLUSTRATED\_BY} relations. Images without
a reliable alignment remain unlinked to disease nodes. To support multimodal retrieval, the images and their associated textual
labels are encoded using the visual and textual branches of
DermLIP~\cite{yan2025derm1m}, respectively. The resulting representations are
indexed to support text-to-image and image-to-image retrieval over the
image nodes.

\Needspace{6\baselineskip}
\subsection{Graph Statistics and Structural Characteristics}
\label{sec:graph_statistics}
The resulting knowledge graph, \textbf{SkinMind-KG}, is a multimodal,
heterogeneous graph containing \(1{,}006{,}079\) nodes and
\(2{,}038{,}419\) edges. The 23 core entity types account for \(99\%\) of all non-image nodes, while the 25 core relation types account for \(99\%\) of all edges excluding \texttt{ILLUSTRATED\_BY}.
The graph includes \(423{,}016\) image nodes, of which \(301{,}561\) are connected
to disease nodes through \texttt{ILLUSTRATED\_BY} relations, providing
explicit associations between clinical images and dermatological concepts.
SkinMind-KG is unweighted and undirected. Its largest connected component
contains \(80.0\%\) of all nodes and \(99.5\%\) of all edges, indicating
that nearly all relational evidence is organized within a shared connected
subgraph. Table~\ref{tab:kg_stats} reports the node and edge counts,
degree distribution, and connected-component statistics. The average
degree is \(4.05\); among non-isolated nodes, the median and 99th-percentile
degrees are \(1\) and \(47\), respectively, while the maximum degree is
\(64{,}345\). This variation in neighborhood size motivates relation-level
probing to guide selective neighbor expansion, enabling the orchestrator to
adapt the breadth of exploration to the local graph structure.

\section{Experiments}
\subsection{Experimental Setting}
\begin{table}[!t]
\caption{Comparison across six dermatology VQA benchmarks (higher is better). Standalone models serve as contextual references and are not directly comparable with the shared SkinMind protocol. Bold and underline mark the best and second-best scores in each column within each of the three blocks. $\Delta$ denotes SkinMind's absolute gain over the best other method in the same policy block, selected independently per column and including Direct Answer. Average is the unweighted mean across six benchmarks; its $\Delta$ uses the highest baseline Average.}
\label{tab:main_results}
\centering
\small
\setlength{\tabcolsep}{4.2pt}
\renewcommand{\arraystretch}{1}
\newcommand{\skinfilledrule}[4]{%
  \arrayrulecolor{#1}\specialrule{\aboverulesep}{0pt}{0pt}%
  \arrayrulecolor{#3}\specialrule{#4}{0pt}{0pt}%
  \arrayrulecolor{#2}\specialrule{\belowrulesep}{0pt}{0pt}%
  \arrayrulecolor{black}%
}
\newcommand{\skinbottomrule}[1]{%
  \arrayrulecolor{#1}\specialrule{\aboverulesep}{0pt}{0pt}%
  \arrayrulecolor{black}\specialrule{\heavyrulewidth}{0pt}{\belowbottomsep}%
}

\resizebox{\textwidth}{!}{%
\begin{tabular}{llccccccc}
\toprule
\textbf{Setting} & \textbf{Model / Method} & \textbf{SkinCon} & \textbf{25Cls} & \textbf{Derm-Edu} & \textbf{DDI} & \textbf{SNU134-4} & \textbf{Morph-R} & \textbf{Average} \\
\skinfilledrule{white}{referenceband}{black}{\lightrulewidth}

\rowcolor{referenceband}
\multicolumn{9}{l}{\textbf{Strong Reference Models}} \\
\rowcolor{referencebg}
 & GPT-5.5~\cite{openai2026gpt55} & \underline{72.01} & \underline{50.40} & \textbf{72.48} & \textbf{65.79} & \textbf{81.50} & \textbf{66.30} & \textbf{68.08} \\
\rowcolor{referencebg}
 & Claude-Sonnet-4.5~\cite{anthropic2025claudesonnet45} & 59.20 & 34.15 & 53.64 & 52.90 & \underline{68.75} & 54.37 & 53.84 \\
\rowcolor{referencebg}
 & DermoGPT-RL~\cite{ru2026dermogpt} & \textbf{76.67} & \textbf{64.60} & \underline{69.68} & \underline{62.80} & 60.00 & \underline{65.48} & \underline{66.54} \\
\rowcolor{referencebg}
 & SkinVL-PubMM~\cite{zeng2025mmskin} & 61.31 & 38.75 & 51.12 & 48.93 & 54.58 & 54.62 & 51.55 \\
\rowcolor{referencebg}
\multirow{-5}{*}{Standalone} & Lingshu-32B~\cite{xu2025lingshu} & 52.39 & 38.40 & 30.29 & 34.91 & 45.83 & 49.55 & 41.90 \\
\skinfilledrule{referencebg}{gptpolicyband}{black!40}{\lightrulewidth}

\rowcolor{gptpolicyband}
\multicolumn{9}{l}{\textbf{GPT-5.4-mini as Policy Model}} \\
\rowcolor{gptpolicybg}
 & Direct Answer & 63.67 & 42.22 & 63.42 & 50.30 & \underline{73.33} & 62.24 & 59.20 \\
\rowcolor{gptpolicybg}
\multirow{-2}{*}{Standard} & Naive RAG~\cite{gao2023naiverag} & \underline{67.25} & \underline{50.40} & 66.79 & 52.90 & 71.25 & \underline{62.50} & 61.85 \\
\skinfilledrule{gptpolicybg}{gptpolicybg}{black!25}{\cmidrulewidth}
\rowcolor{gptpolicybg}
 & GraphRAG~\cite{graphrag} & 64.24 & 46.55 & 65.80 & 54.12 & 71.25 & 60.28 & 60.37 \\
\rowcolor{gptpolicybg}
\multirow{-2}{*}{Graph-based RAG} & LightRAG~\cite{guo2024lightrag} & 64.95 & 45.80 & \underline{68.85} & \underline{58.38} & 72.50 & 62.39 & \underline{62.15} \\
\skinfilledrule{gptpolicybg}{gptpolicybg}{black!25}{\cmidrulewidth}
\rowcolor{gptpolicybg}
 & A-RAG~\cite{du2026arag} & 66.16 & 49.50 & 63.52 & 53.81 & 70.00 & 62.33 & 60.89 \\
\rowcolor{gptpolicybg}
 & RAGentA~\cite{besrour2025ragenta} & 58.63 & 43.85 & 65.58 & 49.70 & 65.83 & 58.71 & 57.05 \\
\rowcolor{gptpolicybg}
\multirow{-3}{*}{Agentic RAG} & \textbf{SkinMind} & \textbf{68.84} & \textbf{54.05} & \textbf{70.60} & \textbf{65.09} & \textbf{85.00} & \textbf{63.84} & \textbf{67.90} \\
\skinfilledrule{gptpolicybg}{gptpolicybg}{black!25}{\cmidrulewidth}
\rowcolor{gptpolicybg}
\multicolumn{2}{r}{\textcolor{skincoral}{\textbf{$\Delta$ vs. Best Baseline}}}
& \textcolor{skincoral}{\textbf{+1.59}} & \textcolor{skincoral}{\textbf{+3.65}} & \textcolor{skincoral}{\textbf{+1.75}} & \textcolor{skincoral}{\textbf{+6.71}} & \textcolor{skincoral}{\textbf{+11.67}} & \textcolor{skincoral}{\textbf{+1.34}} & \textcolor{skincoral}{\textbf{+5.75}} \\
\skinfilledrule{gptpolicybg}{geminipolicyband}{black!40}{\lightrulewidth}

\rowcolor{geminipolicyband}
\multicolumn{9}{l}{\textbf{Gemini-2.5-Flash as Policy Model}} \\
\rowcolor{geminipolicybg}
 & Direct Answer & 66.59 & 47.20 & 66.33 & 59.15 & 65.42 & 58.49 & 60.53 \\
\rowcolor{geminipolicybg}
\multirow{-2}{*}{Standard} & Naive RAG~\cite{gao2023naiverag} & 64.93 & \underline{49.75} & 70.46 & 60.67 & 74.58 & 61.89 & 63.71 \\
\skinfilledrule{geminipolicybg}{geminipolicybg}{black!25}{\cmidrulewidth}
\rowcolor{geminipolicybg}
 & GraphRAG~\cite{graphrag} & 59.64 & 47.65 & 69.79 & 61.13 & 71.67 & 59.71 & 61.60 \\
\rowcolor{geminipolicybg}
\multirow{-2}{*}{Graph-based RAG} & LightRAG~\cite{guo2024lightrag} & \underline{71.28} & 49.05 & 71.92 & \underline{67.07} & 75.00 & 60.62 & \underline{65.82} \\
\skinfilledrule{geminipolicybg}{geminipolicybg}{black!25}{\cmidrulewidth}
\rowcolor{geminipolicybg}
 & A-RAG~\cite{du2026arag} & 70.54 & 46.85 & \underline{72.56} & 62.80 & 72.50 & \underline{62.14} & 64.57 \\
\rowcolor{geminipolicybg}
 & RAGentA~\cite{besrour2025ragenta} & 70.58 & 49.10 & 72.53 & 62.80 & \underline{76.67} & 60.97 & 65.44 \\
\rowcolor{geminipolicybg}
\multirow{-3}{*}{Agentic RAG} & \textbf{SkinMind} & \textbf{75.18} & \textbf{53.03} & \textbf{73.06} & \textbf{67.53} & \textbf{85.11} & \textbf{62.58} & \textbf{69.42} \\
\skinfilledrule{geminipolicybg}{geminipolicybg}{black!25}{\cmidrulewidth}
\rowcolor{geminipolicybg}
\multicolumn{2}{r}{\textcolor{skincoral}{\textbf{$\Delta$ vs. Best Baseline}}}
& \textcolor{skincoral}{\textbf{+3.90}} & \textcolor{skincoral}{\textbf{+3.28}} & \textcolor{skincoral}{\textbf{+0.50}} & \textcolor{skincoral}{\textbf{+0.46}} & \textcolor{skincoral}{\textbf{+8.44}} & \textcolor{skincoral}{\textbf{+0.44}} & \textcolor{skincoral}{\textbf{+3.60}} \\
\skinbottomrule{geminipolicybg}
\end{tabular}%
}
\end{table}

\paragraph{Datasets.}
We evaluate SkinMind on six public dermatology VQA benchmarks spanning
both closed- and open-ended settings. SkinCon~\cite{ru2026dermogpt},
25Cls~\cite{ru2026dermogpt}, Derm-Edu~\cite{ru2026dermogpt},
DDI~\cite{daneshjou2022disparities}, and
SNU134-4~\cite{ru2026dermogpt} are formulated as multiple-choice
benchmarks, whereas Morph-R~\cite{ru2026dermogpt} requires free-form
responses. Collectively, these benchmarks assess morphology recognition,
disease diagnosis, and clinical reasoning across different response
formats.
We use SkinMind-KG as the retrieval corpus. To prevent benchmark leakage
from image overlap, we first remove exact duplicates based on MD5 hashes.
We then exclude retrieved image nodes whose DermLIP cosine similarity to
the query image exceeds \(0.98\), thereby filtering potential near
duplicates. This two-stage procedure is applied identically to all RAG
methods to ensure a fair comparison.

\paragraph{Baselines and Evaluation.}
We organize the baselines into three groups.
(i)~\emph{Standard baselines} comprise zero-shot inference without
retrieval (\emph{Direct Answer}) and conventional single-round
retrieval augmentation (\emph{Naive RAG})~\cite{gao2023naiverag}.
(ii)~\emph{Graph-based RAG methods} include GraphRAG~\cite{graphrag} and LightRAG~\cite{guo2024lightrag}, which leverage graph-structured knowledge for retrieval.
(iii)~\emph{Agentic RAG methods} include A-RAG~\cite{du2026arag} and RAGentA~\cite{besrour2025ragenta}, which interleave model reasoning with iterative retrieval.
To ensure a fair comparison, all methods use the same answering model
and inference settings and receive identical task instructions and
output-format requirements. All retrieval-augmented methods operate on the same SkinMind-KG through a shared retrieval backend. Prompt differences are limited to method-specific retrieval and tool-use instructions. Following the official evaluation protocol of
DermoGPT-RL~\cite{ru2026dermogpt}, we report accuracy on the five
multiple-choice benchmarks and evaluate open-ended responses on Morph-R
using Gemini-2.5-Pro under an LLM-as-a-judge protocol. We use the official
DermoGPT evaluation code without
modification.\footnote{\url{https://github.com/mendicant04/DermoGPT}}
The judge compares each response with its reference answer without
access to the input image. It assesses clinical reasoning, final
diagnosis, and closed-set morphology output, which the official
evaluator combines into the reported Morph-R score.

\paragraph{Implementation Details.}
We instantiate each method with two policy models:
GPT-5.4-mini~\cite{openai2026gpt54mini} and
Gemini-2.5-Flash~\cite{comanici2025gemini25}. Non-agentic methods retrieve the top five textual items and the top five image items, whereas agentic methods allow the policy model to determine
\(k\) dynamically at each retrieval round, with \(k=5\) as the default. Textual and visual retrieval is performed using Qwen3-Embedding-4B~\cite{zhang2025qwen3}
and DermLIP~\cite{yan2025derm1m}, respectively, based on cosine similarity. 
Agentic methods are limited to \(20\) retrieval rounds per question. For generation, we set the temperature to \(1.0\) and the maximum context length to
\(32{,}768\) tokens.
All methods instantiated with the same policy model use identical
generation settings.
The topology of SkinMind-KG is stored in GraphML format, whereas Qdrant stores embedding indices for the shared retrieval backend.

\subsection{Main Results}
\label{sec:main_results}

Table~\ref{tab:main_results} presents the main experimental results across six benchmarks. We highlight three key observations.

\paragraph{SkinMind delivers consistent performance gains.}
SkinMind consistently outperforms the corresponding \emph{Direct Answer}
baseline. With GPT-5.4-mini, it improves the average score from \(59.20\)
to \(67.90\), a gain of \(8.70\) points; with Gemini-2.5-Flash, it
increases the score from \(60.53\) to \(69.42\), a gain of \(8.89\)
points. Most retrieval-augmented methods also improve over direct
answering, supporting the value of SkinMind-KG as an external knowledge
source. Notably, SkinMind outperforms the competing RAG methods under both policy-model settings, indicating that adaptive graph probing enables more effective use of this knowledge.

\paragraph{Adaptive graph probing enables more effective retrieval.}
Compared with LightRAG, SkinMind improves the average score by \(5.75\)
points with GPT-5.4-mini and \(3.60\) points with Gemini-2.5-Flash.
Meanwhile, the gains of other RAG methods vary across benchmarks and
policy models, suggesting that retrieval effectiveness depends not only
on access to external knowledge but also on how that knowledge is
explored. SkinMind adaptively selects entry nodes, probes relations,
and traverses paths to gather query-specific textual and visual
evidence at the required granularity.

\paragraph{SkinMind is competitive with strong standalone reference models.}
Although not directly comparable, standalone models provide useful
reference points. SkinMind with GPT-5.4-mini achieves an average score
of \(67.90\), comparable to GPT-5.5's \(68.08\), while outperforming
DermoGPT-RL, SkinVL-PubMM, and Lingshu-32B by \(1.36\), \(16.35\), and
\(26.00\) points, respectively. This result highlights the effectiveness
of graph-based evidence seeking in narrowing the gap between the
policy model and stronger standalone models.

\subsection{Ablation on Graph Probing Tools}
\label{sec:ablation}
\begin{wraptable}{R}{0.51\textwidth}
\captionsetup{font=small,skip=4pt}
\caption{Ablations with \mbox{GPT-5.4-mini} on DDI and \mbox{SNU134-4}.}
\label{tab:ablation}
\centering
\footnotesize
\setlength{\tabcolsep}{4pt}
\renewcommand{\arraystretch}{1.08}
\begin{tabularx}{\linewidth}{@{}>{\raggedright\arraybackslash}Xr@{\,}rr@{\,}r@{}}
\toprule
\textbf{Method} & \multicolumn{2}{c}{\textbf{DDI}} & \multicolumn{2}{c@{}}{\textbf{SNU134-4}} \\
\midrule
\textcolor{skinteal}{\textbf{SkinMind (Full)}}
& \textbf{65.09} &
& \textbf{85.00} & \\
w/o Graph Exploration
& 62.65 & \textcolor{sectiontext}{($-2.44$)}
& 82.50 & \textcolor{sectiontext}{($-2.50$)} \\
w/o Image Retrieval
& 63.26 & \textcolor{sectiontext}{($-1.83$)}
& 79.17 & \textcolor{sectiontext}{($-5.83$)} \\
w/o Progressive Disclosure
& 64.48 & \textcolor{sectiontext}{($-0.61$)}
& 80.83 & \textcolor{sectiontext}{($-4.17$)} \\
\bottomrule
\end{tabularx}
\end{wraptable}

To examine the roles of graph exploration, image retrieval, and
progressive disclosure, we compare SkinMind with three ablation
variants on DDI and SNU134-4 using GPT-5.4-mini.
\emph{w/o Graph Exploration} disables \texttt{probe\_relations},
\texttt{probe\_neighbors}, and \texttt{probe\_path}, while preserving
textual and visual node retrieval.
\emph{w/o Image Retrieval} disables \texttt{search\_image} and
\texttt{describe\_image}.
In \emph{w/o Progressive Disclosure}, \texttt{read\_node} directly
returns the associated source chunks normally retrieved through a
separate \texttt{read\_chunks} call.

Variants without graph exploration, image retrieval, or progressive
disclosure show lower accuracy on both benchmarks
(Table~\ref{tab:ablation}). On DDI, \emph{w/o Graph Exploration} shows
the largest gap at \(2.44\) points. On SNU134-4,
\emph{w/o Image Retrieval} and \emph{w/o Progressive Disclosure}
show more pronounced gaps of \(5.83\) and \(4.17\) points, respectively.
These patterns are consistent with complementary roles for relational
context, visual references, and selective evidence disclosure in
diagnostic reasoning. Together, these capabilities allow the
orchestrator to adapt both the evidence source and the retrieval depth
to the diagnostic query.

\subsection{Illustrative Case Study}

\begin{figure}[!htbp]
    \centering
    \includegraphics[width=\textwidth]{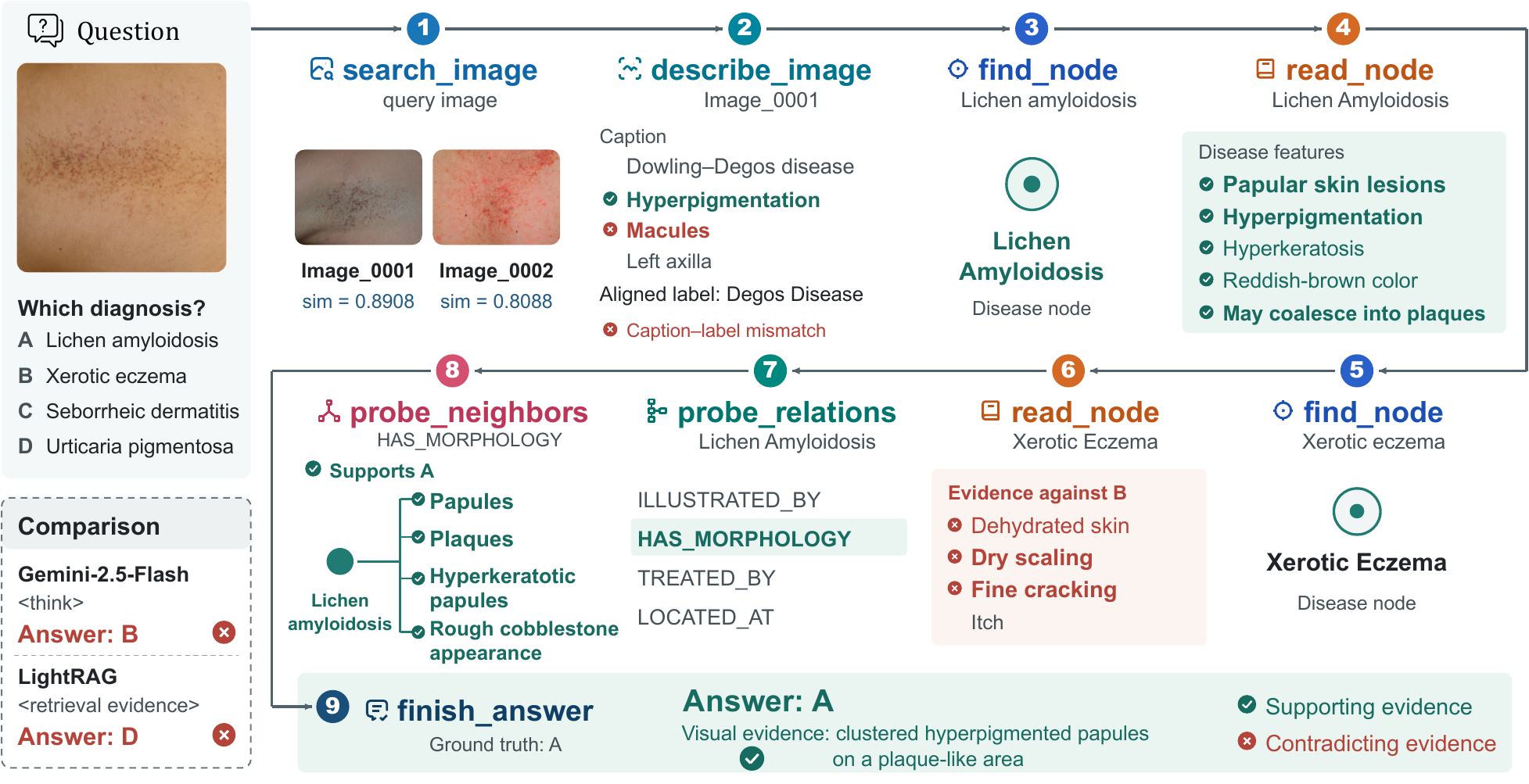}
    \caption{An illustrative graph-probing trajectory of SkinMind. By iteratively retrieving visual and textual evidence and comparing plausible diagnoses, SkinMind correctly identifies lichen amyloidosis, whereas Gemini-2.5-Flash and LightRAG produce incorrect predictions.}
    \label{fig:case_study}
    \Description{A dermatology diagnosis example showing a clinical image, candidate answers, successive graph-probing tool calls, retrieved evidence, and the final answer.}
\end{figure}

Figure~\ref{fig:case_study} illustrates a reasoning trajectory of SkinMind for dermatological diagnosis. Given a clinical image and a multiple-choice question, SkinMind progressively retrieves visually similar cases, extracts dermatological concepts, and explores relevant disease nodes to gather supporting evidence and eliminate competing diagnoses. By integrating the observed reddish-brown hyperpigmented papules and macules, coalescing lesions, and signs of chronic scratching with the retrieved knowledge, SkinMind correctly identifies the condition as lichen amyloidosis. In contrast, Gemini-2.5-Flash and
LightRAG select incorrect answers. This example illustrates how
SkinMind's iterative evidence-seeking process supports fine-grained,
evidence-grounded clinical reasoning.

\subsection{Context Efficiency Analysis}
\label{subsec:token_efficiency}

\begin{wrapfigure}{r}{0.49\textwidth}
  \centering
    \captionsetup{font=small,skip=4pt}
  \includegraphics[width=\linewidth,trim=0 10bp 0 0,clip]{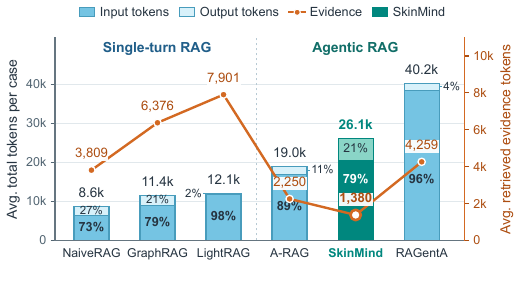}
  \caption{Token-efficiency comparison across RAG methods on Morph-R using Gemini-2.5-Flash as the policy model. Stacked bars show average input and output tokens, while the line indicates average retrieved evidence tokens. SkinMind retrieves the most compact evidence set.}
  \label{fig:token_comparison}
    \Description{Stacked bars show average input and output tokens on Morph-R for six retrieval-augmented methods; a line shows average retrieved evidence tokens on a second axis.}
\end{wrapfigure}

Figure~\ref{fig:token_comparison} compares the average token consumption of
SkinMind against five representative retrieval-augmented baselines on the Morph-R benchmark. The stacked bars show input and output tokens
(left axis), while the line reports the average number of retrieved
evidence tokens in the model context (right axis).

\noindent\textbf{Evidence efficiency.}
Most notably, SkinMind achieves the strongest task performance while
using the fewest evidence tokens among all evaluated methods, averaging
\(1{,}380\) tokens per question. By comparison, LightRAG uses
\(7{,}901\) evidence tokens, \(5.7\times\) as many as SkinMind. This
pattern suggests that SkinMind's gains are associated with more targeted
evidence selection rather than a larger retrieved context. Its
hierarchical graph-probing interfaces support selective access to
relevant textual and visual evidence at different levels of granularity.

\Needspace{18\baselineskip}

\noindent\textbf{Reasoning-oriented token allocation.}
SkinMind allocates \(21\%\) of its token usage to model outputs,
amounting to \(5{,}576\) tokens per question. These outputs include both
the final answer and the intermediate decisions involved in navigating
the graph and selecting evidence. In comparison, LightRAG allocates
\(2\%\) of its token usage (\(291\) tokens) to model outputs, with
\(11{,}802\) input tokens consumed through upfront context construction.
Aggregated across calls, total usage is approximately \(26.1\mathrm{k}\)
tokens for SkinMind and \(12.1\mathrm{k}\) for LightRAG. SkinMind's
multi-round inputs include the interaction history and tool instructions
alongside retrieved evidence. This allocation characterizes its
iterative evidence-acquisition process: the model progressively selects
what to retrieve and uses the resulting observations to guide subsequent
graph-probing actions.

\section{Conclusion}
We introduced \textbf{SkinMind}, a multimodal agentic RAG framework that
interactively probes a dermatology knowledge graph for textual, relational,
and visual evidence. Equipped with hierarchical graph-probing tools,
SkinMind interleaves reasoning with adaptive exploration of textual,
relational, and visual evidence, progressively constructing query-specific
evidence subgraphs. We also developed \textbf{SkinMind-KG} to provide a
comprehensive and traceable foundation of domain-specific knowledge.
Across six dermatology VQA benchmarks and two policy
models, SkinMind consistently outperforms direct answering and competing
RAG methods while retrieving more targeted evidence. These results support adaptive graph probing as a promising approach to evidence-grounded dermatology reasoning.

\FloatBarrier
\bibliographystyle{plainnat}
\bibliography{references}

\clearpage
\appendix
\renewcommand{\runningtitle}{SkinMind: Appendix}
\fancyhead[L]{\nouppercase{\small\textbf{\runningtitle}}}
\setcounter{figure}{0}
\setcounter{table}{0}
\setcounter{equation}{0}
\renewcommand{\thefigure}{S\arabic{figure}}
\renewcommand{\thetable}{S\arabic{table}}
\renewcommand{\theequation}{S\arabic{equation}}
\renewcommand{\theHfigure}{appendix.\arabic{figure}}
\renewcommand{\theHtable}{appendix.\arabic{table}}
\renewcommand{\theHequation}{appendix.\arabic{equation}}
\section*{Appendix}
\addcontentsline{toc}{section}{Appendix}
\section{SkinMind-KG Data Appendix}
\label{sec:supp-kg}

This section documents the data-processing and graph-construction details that
are abbreviated in the main paper. SkinMind-KG combines source-linked textual
knowledge with image nodes under a dermatology-specific schema. The
construction pipeline preserves the source chunk identifiers associated with
extracted entities and relations, thereby enabling retrieved graph content to
be traced to its supporting text.

\subsection{Source Composition and Provenance}

The textual component of SkinMind-KG is constructed from 769 documents,
including 227 clinical guidelines and 542 textbooks published since 2015.
The multimodal component contains 423,016 images from
Derm1M~\cite{yan2025derm1m}.

\subsection{Text Parsing, Cleaning, and Chunking}

We parse each source document into page-level structured text using MinerU.
Rule-based filtering removes non-textual elements, including images and
tables, while incomplete adjacent paragraphs are merged. The processed text
is then segmented into 1,200-token chunks with a 100-token overlap for entity
and relation extraction. Each chunk retains its document-level provenance and
serves as the source unit associated with the extracted entities and
relations.

\subsection{Disease Ontology and Extraction Schema}

We adopt the PanDerm ontology~\cite{yan2025panderm} as the taxonomic backbone
of SkinMind-KG. It contains 127 dermatological disease concepts and their
curated hierarchical relationships. We represent these concepts as
\texttt{Disease} and \texttt{DiseaseCategory} nodes and encode their hierarchy
using \texttt{IS\_SUBTYPE\_OF} relations. The resulting ontology is preloaded
into SkinMind-KG before knowledge extraction. In consultation with senior
dermatologists, we further define the 23 entity types in
Table~\ref{tab:supp-entity-schema} and the 25 relation types in
Table~\ref{tab:supp-relation-schema}.

Entities and relations are extracted from each text chunk using the system and
user prompt templates reproduced in
Section~\ref{sec:supp-extraction-prompt}.
Relations are extracted using canonical source--target roles, whereas the final
graph is undirected and unweighted.

\begin{table}[!t]
\caption{Entity schema for constructing the textual subgraph of
    SkinMind-KG. The schema contains 23 entity types; group labels organize
    these types but are not extracted as entities.}
    \label{tab:supp-entity-schema}
    \centering
    \small
    \begin{tabularx}{\textwidth}{@{}p{0.15\textwidth} p{0.23\textwidth} Y@{}}
        \toprule
        \textbf{Schema group} & \textbf{Entity type} & \textbf{Definition} \\
        \midrule
        Disease taxonomy & Disease & A specific named dermatologic disease or diagnosis; generic disease states and risk-tier wrappers are excluded. \\
        & DiseaseCategory & A dermatologic category used for disease taxonomy. \\
        & Severity & An explicitly stated severity, grade, stage, score, risk tier, or prognosis descriptor. \\
        \addlinespace
        Clinical presentation & Symptom & A subjective symptom or naked-eye clinical sign. \\
        & Morphology & A lesion shape, structure, border, or configuration. \\
        & Color & A color or pigmentary feature. \\
        & AnatomicalSite & A body site or anatomical location. \\
        & DistributionPattern & A lesion distribution pattern. \\
        \addlinespace
        Etiology and risk & Etiology & A non-infectious cause, pathogenesis, or mechanism. \\
        & Pathogen & An infectious organism or clinically meaningful pathogen category. \\
        & Trigger & A precipitating or aggravating factor. \\
        & RiskFactor & A factor associated with disease likelihood. \\
        & Population & A patient or demographic group. \\
        \addlinespace
        Diagnostic assessment & Test & A diagnostic test, examination, or procedure. \\
        & TestFinding & A finding identified by a test, instrument, examination maneuver, or procedure. \\
        & HistopathologicFinding & A microscopic pathological feature identified through tissue examination. \\
        & DifferentialFeature & A feature explicitly used to distinguish one diagnosis from another. \\
        \addlinespace
        Therapeutic management & Treatment & A treatment strategy, modality, or procedure. \\
        & Drug & A specific medication or drug. \\
        & DrugClass & A pharmacological drug class. \\
        & Contraindication & A condition or situation in which a drug or treatment should be avoided. \\
        \addlinespace
        Clinical outcomes & Complication & A consequence of a disease or treatment. \\
        Terminology & Alias & An explicit synonym, abbreviation, acronym, or alternative name. \\
        \bottomrule
    \end{tabularx}
    
\end{table}

\begin{table}[p]
\caption{Relation schema for constructing the textual subgraph of
    SkinMind-KG. Arrows specify canonical source-to-target roles during
    extraction; the resulting graph is undirected and unweighted.}
    \label{tab:supp-relation-schema}
    \centering
    \small
    \renewcommand{\arraystretch}{0.94}
    \begin{tabularx}{\textwidth}{@{}p{0.22\textwidth} p{0.16\textwidth} Y@{}}
        \toprule
        \textbf{Relation type} & \textbf{Meaning} & \textbf{Schema-defined endpoint roles} \\
        \midrule
        \texttt{IS\_SUBTYPE\_OF} & Taxonomic parent & Disease$\rightarrow$Disease/DiseaseCategory; DiseaseCategory$\rightarrow$DiseaseCategory; Drug$\rightarrow$Drug; DrugClass$\rightarrow$DrugClass; Treatment$\rightarrow$Treatment; Test$\rightarrow$Test; Pathogen$\rightarrow$Pathogen; Symptom$\rightarrow$Symptom \\
        \texttt{HAS\_SEVERITY} & Severity or stage & Disease$\rightarrow$Severity \\
        \texttt{HAS\_SYMPTOM} & Clinical sign or symptom & Disease$\rightarrow$Symptom \\
        \texttt{HAS\_MORPHOLOGY} & Lesion morphology & Disease$\rightarrow$Morphology \\
        \texttt{HAS\_COLOR} & Color feature & Disease$\rightarrow$Color \\
        \texttt{LOCATED\_AT} & Anatomical involvement & Disease$\rightarrow$AnatomicalSite \\
        \texttt{HAS\_DISTRIBUTION} & Distribution pattern & Disease$\rightarrow$DistributionPattern \\
        \texttt{HAS\_ETIOLOGY} & Non-infectious etiology & Disease$\rightarrow$Etiology \\
        \texttt{CAUSED\_BY} & Infectious cause & Disease$\rightarrow$Pathogen \\
        \texttt{TRIGGERED\_BY} & Precipitating factor & Disease$\rightarrow$Trigger \\
        \texttt{HAS\_RISK\_FACTOR} & Predisposing factor & Disease$\rightarrow$RiskFactor \\
        \texttt{COMMON\_IN} & Associated population & Disease$\rightarrow$Population \\
        \texttt{DIAGNOSED\_BY} & Diagnostic examination & Disease$\rightarrow$Test \\
        \texttt{HAS\_TEST\_FINDING} & Test-derived finding & Disease$\rightarrow$TestFinding \\
        \texttt{FOUND\_BY} & Finding-to-test link & TestFinding/HistopathologicFinding$\rightarrow$Test \\
        \texttt{HAS\_HISTOPATHOLOGY} & Microscopic feature & Disease$\rightarrow$HistopathologicFinding \\
        \texttt{DIFFERENTIAL\_WITH} & Differential diagnosis & Disease$\rightarrow$Disease \\
        \texttt{DISTINGUISHED\_BY} & Discriminative feature & Disease$\rightarrow$DifferentialFeature \\
        \texttt{TREATED\_BY} & Treatment modality & Disease$\rightarrow$Treatment \\
        \texttt{USES\_DRUG} & Specific medication & Disease/Treatment$\rightarrow$Drug \\
        \texttt{USES\_DRUG\_CLASS} & Medication class & Disease$\rightarrow$DrugClass \\
        \texttt{BELONGS\_TO\_DRUG\_CLASS} & Class membership & Drug$\rightarrow$DrugClass \\
        \texttt{CONTRAINDICATED\_WITH} & Contraindicated condition & Drug/Treatment$\rightarrow$Contraindication \\
        \texttt{HAS\_COMPLICATION} & Clinical outcome & Disease$\rightarrow$Complication \\
        \texttt{HAS\_ALIAS} & Synonym or abbreviation & Disease/Drug/DrugClass/Test/Pathogen/Treatment$\rightarrow$Alias \\
        \bottomrule
    \end{tabularx}
    
\end{table}

\subsection{Merging, Deduplication, and Conflict Resolution}

After extraction, records with the same entity name are grouped, duplicate
source identifiers are removed, and relations connecting the same pair of
nodes under the same relation type are consolidated. Self-loops are discarded.
The grouping step is deterministic; an LLM is then used within each group to
synthesize the collected descriptions and detect genuine factual conflicts.
As specified by the prompt in
Section~\ref{sec:supp-description-merge-prompt}, two claims are considered
conflicting only when their subject, attribute, clinical scope, metric, and
unit match. Otherwise, their information is treated as complementary and
integrated into a single description.

Integrating heterogeneous sources may introduce contradictory claims about
the same entity or relation. When a conflict is detected, the synthesis prompt
preserves the opposing claims and generates targeted search queries without
resolving the disagreement. Gemini-3.1-Pro subsequently uses these queries to
retrieve relevant evidence through Google Search and determine whether the
conflict can be resolved. This procedure processed 1,375 description-level
conflicts, automatically resolving 1,366 (99.3\%) and referring the remaining
nine cases for expert review.

\subsection{Image-Node Integration}

Each image is represented as a separate \texttt{Image} node with a stable
\texttt{img\_<hash>} identifier. Its payload can include the filename,
caption, body location, symptom labels, skin concepts, age, gender, raw
diagnosis labels, aligned diseases, source, and source type. During graph
construction, the \texttt{aligned\_diseases} labels are mapped to existing
\texttt{Disease} nodes through a three-stage procedure. First, normalized
labels are matched exactly against disease-node names. Second, for each
unmatched label, Qwen3-Embedding-4B retrieves the 10 semantically closest
disease nodes. Third, GPT-5.4-mini disambiguates among these candidates and
determines whether a reliable alignment can be established. Each successful
alignment connects the disease and image nodes through an
\texttt{ILLUSTRATED\_BY} relation; images without a reliable alignment remain
unlinked. The graph contains 608,420 such edges; because an image can align to
more than one disease, this count exceeds the number of image nodes. Of
423,016 images, 301,561 are linked to at least one disease and 121,455 remain
unlinked.

For image-conditioned retrieval, the query image is encoded by the DermLIP
visual tower as a 512-dimensional L2-normalized vector. Qdrant returns images
with cosine similarity of at least 0.8. Matches above 0.98 are excluded as
potential duplicates.

\subsection{Graph Storage and Representative Records}

SkinMind-KG uses three complementary storage layers. GraphML stores nodes and
edges, key--value stores retain text chunks and entity-to-chunk mappings, and
Qdrant indexes embeddings for entities, relations, text chunks, captions, and
images.
Entity and textual-relation records retain \texttt{source\_id} values that
link them to source chunks, while multimodal links retain image-source
identifiers.

\begin{lstlisting}[basicstyle=\ttfamily\footnotesize,captionpos=t,aboveskip=6pt,belowskip=6pt,caption={Representative node, relation, source-chunk, and image-link records stored in SkinMind-KG.},label={lst:supp-records}]
NODE
{"entity_id":"Basal Cell Carcinoma",
 "entity_type":"Disease",
 "description":"Basal Cell Carcinoma is ...",
 "source_id":"chunk-4400cb15...<SEP>chunk-a780475c...",
 "Conflict_Status":"None"}

EDGE
{"src":"Psoriasis","tgt":"Topical Corticosteroids",
 "keywords":"USES_DRUG_CLASS",
 "description":"Mild psoriasis is treated with topical corticosteroids.",
 "source_id":"chunk-xxxx"}

SOURCE CHUNK
{"chunk-4400cb15...":{"content":"Basal cell carcinoma ...",
 "full_doc_id":"doc-be8ac60b...","tokens":1187,
 "chunk_order_index":3}}

IMAGE NODE AND LINK
{"entity_id":"img_d59f310376da370c","entity_type":"Image",
 "image_filename":"IIYI/26549_2.png","image_age":"15",
 "image_source":"IIYI_chinese","image_source_type":"forum"}
{"src":"Epidermolysis Bullosa","tgt":"img_d59f310376da370c",
 "keywords":"ILLUSTRATED_BY",
 "source_id":"mm_img:IIYI_chinese"}
\end{lstlisting}

\section{Experimental Reproducibility}
\label{sec:supp-reproducibility}

\subsection{Hardware and Software Environment}

Experiments were run on eight NVIDIA A800-SXM4 GPUs with 80~GB of memory per
GPU. The environment used CPython 3.10.15, PyTorch 2.5.1, CUDA 12.4
(\texttt{cu124}), and cuDNN 9.1.0.

Qdrant v1.17.0 and \texttt{qdrant-client} 1.18.0 were used for vector
indexing and retrieval. Textual entities, relations, chunks, and image
captions share a 2,560-dimensional Qwen3-Embedding-4B cosine space, while
DermLIP uses a separate 512-dimensional joint text--image cosine space.
DermoGPT-RL was served locally for multimodal inference with vLLM 0.7.3.
Hosted policy and judge models were accessed through an OpenAI-compatible
interface.

\subsection{Comparative Method Implementation Details}

All retrieval baselines share the same corpus and Qdrant chunk collection.
They use Qwen3-Embedding-4B (2,560 dimensions) for text retrieval and
DermLIP (512 dimensions) for image-similarity retrieval. Each method is
evaluated with Gemini-2.5-Flash and GPT-5.4-mini through the same
OpenAI-compatible interface. Table~\ref{tab:supp-retrieval-parameters}
summarizes the configurations. Retrieval counts apply per modality:
five text or graph results plus five images give 10 results in total.

\begin{table}[!htbp]
\caption{Retrieval configurations for the evaluated methods. Columns
    specify the retrieval budget, text and image encoders, retrieval
    procedure, and handling of retrieved context.}
    \label{tab:supp-retrieval-parameters}
    \centering
    \small
    \renewcommand{\arraystretch}{0.94}
    \begin{tabularx}{\textwidth}{@{}p{0.12\textwidth} p{0.17\textwidth} p{0.18\textwidth} Y Y@{}}
        \toprule
        \textbf{Method} & \textbf{Retrieval budget} & \textbf{Retrieval encoders} &
        \textbf{Retrieval procedure} & \textbf{Context handling} \\
        \midrule
        Direct Answer & None & None & No retrieval & No retrieved context \\
        Naive RAG & 5 text chunks; 5 images & Qwen3-Embedding-4B, 2,560d; DermLIP, 512d & One dense text search and one image-similarity search & All retrieved chunks and images are passed to the generator \\
        GraphRAG & 10 entities; 10 relations; 5 images & Qwen3-Embedding-4B, 2,560d; DermLIP, 512d & Local graph search over entities and relations, plus image-similarity retrieval & Retrieved entities, relations, source text, and images are passed to the generator \\
        LightRAG & Entity/relation top-$k=10$; 5 chunks; 5 images & Qwen3-Embedding-4B, 2,560d; DermLIP, 512d & Joint entity, relation, and chunk retrieval, plus image-similarity retrieval & Retrieved chunks, graph records, and images are combined in the generator context \\
        A-RAG & Text top-$k$: default 5, maximum 20; 5 images; 5 sentence matches per chunk & Qwen3-Embedding-4B, 2,560d; DermLIP, 512d & Agent-selected keyword, semantic, full-chunk, and image retrieval & Retrieved evidence is accumulated across tool calls \\
        RAGentA & Follow-up text top-$k=10$; 5 images & Qwen3-Embedding-4B, 2,560d; DermLIP, 512d & Iterative dense--sparse text retrieval plus image-similarity retrieval & Evidence is accumulated across follow-up retrieval rounds \\
        SkinMind & Policy-selected per tool call; default 5 text/graph items and 5 images & Qwen3-Embedding-4B, 2,560d; DermLIP, 512d & Policy-controlled node search, graph probing, source-text reading, and image retrieval & Tool observations are accumulated and compressed during inference \\
        \bottomrule
    \end{tabularx}
    
\end{table}

\clearpage
\section{Failure Case Analysis}
\label{sec:supp-error-analysis}
\enlargethispage{\baselineskip}

We also analyze two failure cases of SkinMind to examine its diagnostic
reasoning and identify opportunities for more targeted evidence gathering.
Both cases are drawn from the 25Cls benchmark~\cite{ru2026dermogpt}, with
Gemini-2.5-Flash~\cite{comanici2025gemini25} as the policy model.
Figure~\ref{fig:supp-error-overlap} shows a basal cell
carcinoma (BCC) case predicted as keratoacanthoma (KA), while
Figure~\ref{fig:supp-error-focused} shows pilomatricoma predicted as
cellular neurothekeoma. In both cases, SkinMind links retrieved disease
knowledge to the lesion's appearance. The first leaves
growth history and central keratin unverified, while exploration in the
second centers on the favored candidate. These observations suggest that
the policy model's agentic tool-use strategy could be further improved to
gather evidence that better distinguishes competing diagnoses. Future work
could explore tool-use training for agentic RAG to encourage more targeted
tool selection and comparative exploration.

\noindent\begin{minipage}{\textwidth}
    \centering
    \captionsetup{font=small,skip=3pt}
    \includegraphics[width=0.88\textwidth]{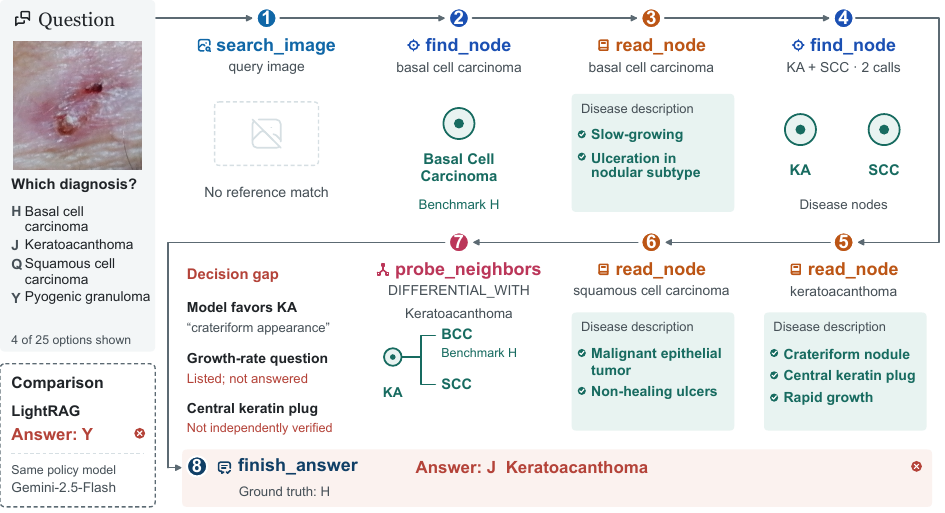}
    \captionof{figure}{Candidate differentiation after retrieval. Four of 25 options
    are shown. Numbered groups reorganize logged calls; red text marks
    retrospective analysis. SCC denotes squamous cell carcinoma.}
    \label{fig:supp-error-overlap}
    \Description{The original query image and four selected options accompany
    image search, disease-node searches and reads, and a differential
    neighbor probe. Retrieved descriptions include BCC's slow growth and
    KA's rapid growth and central keratin plug. The model favors KA's
    crateriform appearance, but its growth-rate question remains unanswered.}
\end{minipage}
\par\vspace{1pt}
\noindent\begin{minipage}{\textwidth}
    \centering
    \captionsetup{font=small,skip=3pt}
    \includegraphics[width=0.88\textwidth]{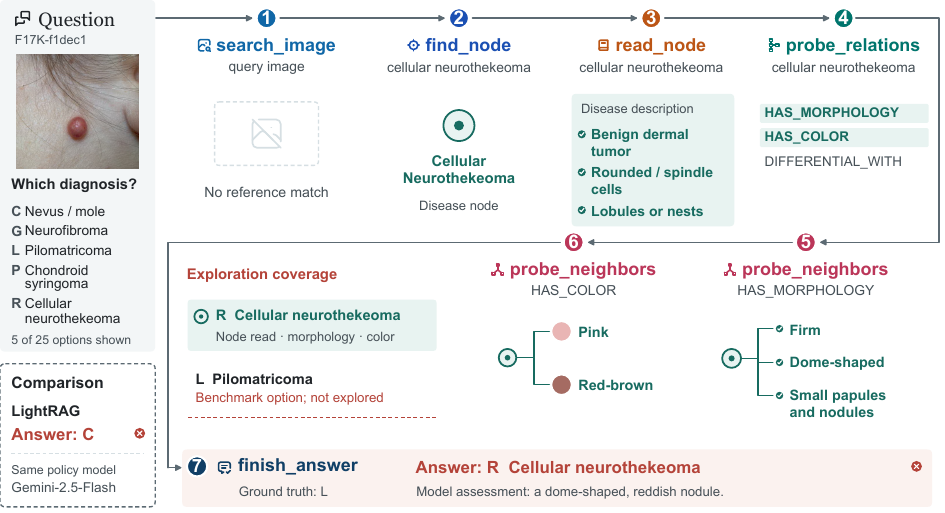}
    \captionof{figure}{Candidate-focused exploration. Five of 25 options are shown;
    morphology and color probes concentrate on cellular neurothekeoma.}
    \label{fig:supp-error-focused}
    \Description{A query nodule is paired with node reading, relation
    discovery, and two neighbor probes for cellular neurothekeoma.
    Returned features include dome-shaped, small papules and nodules,
    pink, and red-brown. The final answer is R; the benchmark label is L.}
\end{minipage}

\clearpage
\section{Prompts and Tool Interfaces}
\label{sec:supp-extraction-prompt}

For reproducibility, this section reports the prompts and tool instructions
used for knowledge-graph construction and SkinMind inference. During entity
and relation extraction,
\texttt{\{language\}} is set to \texttt{English},
\texttt{\{tuple\_delimiter\}} to \texttt{<|\#|>}, and
\texttt{\{completion\_delimiter\}} to \texttt{<|COMPLETE|>}. The
\texttt{\{entity\_ontology\}} and
\texttt{\{relationship\_ontology\}} placeholders are filled with the
corresponding blocks shown below; \texttt{\{input\_text\}} is replaced by
the current source chunk. The injected few-shot example is omitted for brevity.
The PanDerm hierarchy is preloaded into the graph, while the
\texttt{\{prior\_ontology\}} prompt placeholder is empty.
For LaTeX compatibility, typographic dashes and arrows are rendered as their
ASCII equivalents; the wording is otherwise unchanged.

\subsection{Knowledge-Extraction System Prompt}

\begin{lstlisting}
---Role---
You are a Dermatology Knowledge Graph Specialist responsible for extracting dermatology-specific entities and relationships from the input text.

{entity_ontology}

{relationship_ontology}

{prior_ontology}

---Core Principles---

* The entire output MUST be written entirely in `{language}`. This includes ALL `entity_name`, `entity_type`, `entity_description`, `relationship_keywords`, and `relationship_description` fields. Translate every term -- including disease names, drug names, anatomical sites, symptoms, pathogens, and any other domain-specific proper nouns -- into `{language}` using the most widely accepted standardized terminology (e.g., for medical content, use standard English medical terminology such as ICD / MeSH / SNOMED-style names). If the source text is in another language (e.g., Chinese), you MUST translate every entity and relationship into `{language}`. The only exceptions are: fixed schema labels (`entity`, `relation`), the `{tuple_delimiter}` / `{completion_delimiter}` markers, and well-established Latin scientific names or universally untranslatable abbreviations (e.g., `BCC`, `HPV`, `EGFR`).
* Extract only entities and relationships that are explicitly stated or directly supported by the input text. Do not infer from general medical knowledge.
* Do not extract negated findings as positive entities or relationships. For uncertain statements (e.g., `may occur`), extract only when clinically meaningful and directly supported. Preserve conditional context in descriptions.
* Write descriptions as encyclopedic facts about the entity itself, not as a summary of the source. Forbidden meta-narrative includes references like `this article / paper / text / guideline / document / our study`, and document-structure phrases like `discussed in / mentioned in / referenced in / covered by / described in`. State clinical facts directly (e.g., "BCC is treated with surgical excision" -- NOT "BCC is discussed in surgical treatment pathways"). Do not add unsupported information.

---Instructions---

1. Entity Extraction

* Entity Details: For each identified entity, extract:
    * `entity_name`: Use a standardized medical term when the text clearly refers to one. Title Case. Ensure consistent naming throughout.
    * `entity_type`: One of `{entity_types}`. Use exact casing as listed. Do not create new types. If none apply, skip the entity.
    * `entity_description`: A comprehensive description of the entity's attributes, clinical meaning, or role, based solely on the input text.

* Entity Type Definitions:
    * `Disease`: A specific, named dermatologic disease or diagnosis (e.g., `Psoriasis`, `Melanoma`, `Atopic Dermatitis`). Do NOT extract generic disease-state descriptors used to characterize a patient population or trial-eligibility setting (such as `Unresectable Disease`, `Advanced Disease`, `Metastatic Disease`, `Resectable Disease`, `Locally Advanced Disease`, `Recurrent Disease`, `Refractory Disease`); these are clinical states, not diseases themselves. Also do NOT extract risk-tier wrapper phrases like `Low-Risk Tumors`, `High-Risk Lesions` -- decompose into Disease + Severity.
    * `DiseaseCategory`: Dermatologic disease category used for taxonomy.
    * `Severity`: Disease severity, grade, stage, or score descriptor exactly as stated. Examples: `Mild`, `Stage IV`, `High-Risk`, `Low-Risk`, `Poor Prognosis`. Do NOT extract entities like `Risk Status`, `Risk Category`, `Severity Level`.
    * `Symptom`: Subjective symptom or naked-eye clinical sign.
    * `Morphology`: Lesion shape, structure, border, or configuration.
    * `Color`: Color or pigmentary feature.
    * `AnatomicalSite`: Body site or anatomical location.
    * `DistributionPattern`: Lesion distribution pattern.
    * `Etiology`: Non-infectious cause, pathogenesis, or mechanism.
    * `Pathogen`: Infectious organism or clinically meaningful pathogen category.
    * `Trigger`: Precipitating or aggravating factor.
    * `RiskFactor`: Predisposing factor associated with disease likelihood.
    * `Population`: Patient group or demographic group.
    * `Test`: Diagnostic test, examination, or procedure.
    * `TestFinding`: Finding identified by test, instrument, examination maneuver, or procedure.
    * `HistopathologicFinding`: Microscopic pathological feature identified through tissue examination.
    * `DifferentialFeature`: Feature explicitly used to distinguish one diagnosis from another.
    * `Treatment`: Treatment strategy, modality, or procedure.
    * `Drug`: Specific medication or drug.
    * `DrugClass`: Pharmacological drug class.
    * `Contraindication`: Condition or situation in which a drug or treatment should be avoided.
    * `Complication`: Disease or treatment consequence.
    * `Alias`: Explicit synonym, abbreviation, acronym, or alternative name.

* Exclusion Rules -- DO NOT extract the following as entities:
    * Numeric values with units (e.g., `1 cm`, `5 mg`, `10%`, `240-mg`).
    * Research metadata: study names, trial names, journal names, conference names, citations.
    * Document metadata: years, dates, version numbers, chapters, figures, tables.
    * Overly generic categories (e.g., `Supportive Care`, `Symptomatic Treatment`).
    * Non-standard descriptive phrases formed by a vague qualifier (`optimal`, `appropriate`, `selected`, `prior`, `routine`, `standard`) + a generic noun, when they do not name a recognized clinical concept (e.g., skip `Optimal Surgery`, `Appropriate Therapy`; keep `Mohs Surgery`, `Wide Local Excision`).
    * Isolated entities: drop any entity that neither participates in an Allowed Relationship nor carries standalone value (e.g., a recognized alias or named drug). Aim for a connected graph, not a glossary.
    * Staging or severity combined with disease names -- extract them separately and link via `HAS_SEVERITY`.
    * Trial-eligibility or treatment-context descriptors (e.g., `Unresectable / Advanced / Metastatic / Recurrent / Refractory / Progressive / Stable Disease`) when the source text uses them as a patient setting or treatment context (typical pattern: "in patients with X, drug Y is given") rather than as a clinically defined disease entity in its own right. Judge by semantic role, not wording -- skip both the entity and any relationship to it.
    * Abstract meta-concepts naming a CLASSIFICATION DIMENSION rather than a clinical entity (e.g., `Risk Status`, `Risk Category`, `Prognostic Category`, `Treatment Approach`, `Management Strategy`). Extract the concrete instances under them, not the dimension itself.
    * Risk-tier or prognosis-tier wrapper phrases (e.g., `Low-Risk Tumors`, `High-Risk Lesions`, `Poor Prognosis Features`). Decompose into the named disease + a Severity tag (`Low-Risk`, `High-Risk`, `Poor Prognosis`) linked via `HAS_SEVERITY`. Do NOT keep the wrapper as a standalone entity.

* Extraction Granularity:
    * Split composite clinical descriptions into distinct entities. For example, `well-demarcated erythematous plaques with silvery scales` -> separate Morphology, Color, Morphology, and Symptom entities.
    * When an abbreviation or synonym is explicitly defined, extract both the canonical entity and the abbreviation as `Alias`, then create a `HAS_ALIAS` relationship.

2. Relationship Extraction

* Allowed Relationship Schema -- extract ONLY the following directed relationships:

  Taxonomy:
    * Disease -> Disease: `IS_SUBTYPE_OF`, `DIFFERENTIAL_WITH`
    * Disease -> DiseaseCategory: `IS_SUBTYPE_OF`
    * DiseaseCategory -> DiseaseCategory: `IS_SUBTYPE_OF`
    * Drug -> Drug: `IS_SUBTYPE_OF`
    * DrugClass -> DrugClass: `IS_SUBTYPE_OF`
    * Treatment -> Treatment: `IS_SUBTYPE_OF`
    * Test -> Test: `IS_SUBTYPE_OF`
    * Pathogen -> Pathogen: `IS_SUBTYPE_OF`
    * Symptom -> Symptom: `IS_SUBTYPE_OF`

  Clinical Presentation:
    * Disease -> Severity: `HAS_SEVERITY`
    * Disease -> Symptom: `HAS_SYMPTOM`
    * Disease -> Morphology: `HAS_MORPHOLOGY`
    * Disease -> Color: `HAS_COLOR`
    * Disease -> AnatomicalSite: `LOCATED_AT`
    * Disease -> DistributionPattern: `HAS_DISTRIBUTION`

  Etiology & Risk:
    * Disease -> Etiology: `HAS_ETIOLOGY`
    * Disease -> Pathogen: `CAUSED_BY`
    * Disease -> Trigger: `TRIGGERED_BY`
    * Disease -> RiskFactor: `HAS_RISK_FACTOR`
    * Disease -> Population: `COMMON_IN`

  Diagnosis:
    * Disease -> Test: `DIAGNOSED_BY`
    * Disease -> TestFinding: `HAS_TEST_FINDING`
    * TestFinding -> Test: `FOUND_BY`
    * Disease -> HistopathologicFinding: `HAS_HISTOPATHOLOGY`
    * HistopathologicFinding -> Test: `FOUND_BY`
    * Disease -> DifferentialFeature: `DISTINGUISHED_BY`

  Treatment:
    * Disease -> Treatment: `TREATED_BY`
    * Disease -> Drug: `USES_DRUG`
    * Disease -> DrugClass: `USES_DRUG_CLASS`
    * Treatment -> Drug: `USES_DRUG`
    * Drug -> DrugClass: `BELONGS_TO_DRUG_CLASS`
    * Drug -> Contraindication: `CONTRAINDICATED_WITH`
    * Treatment -> Contraindication: `CONTRAINDICATED_WITH`

  Outcome:
    * Disease -> Complication: `HAS_COMPLICATION`

  Terminology:
    * Disease -> Alias: `HAS_ALIAS`
    * Drug -> Alias: `HAS_ALIAS`
    * DrugClass -> Alias: `HAS_ALIAS`
    * Test -> Alias: `HAS_ALIAS`
    * Pathogen -> Alias: `HAS_ALIAS`
    * Treatment -> Alias: `HAS_ALIAS`

* Relationship Details: For each relationship, extract:
    * `source_entity`: Must exactly match an extracted entity name.
    * `target_entity`: Must exactly match an extracted entity name.
    * `relationship_keywords`: First keyword must be the exact relationship type. Additional keywords separated by comma `,`.
    * `relationship_description`: A concise explanation based solely on the input text.

* All relationships are directed -- use the exact source->target direction defined above. Output each unique relationship once with a comprehensive description. Do not create relationships between incompatible entity types.
* If one statement contains multiple targets, decompose into multiple binary relationships.

* **CRITICAL -- Endpoint Closure Rule**: Both `source_entity` and `target_entity` of every relationship MUST have already been extracted as entities in the entity list above, with the EXACT same name and a valid `entity_type`.

3. Output Format

* Output all entities first, then all relationships. Prioritize relationships central to the clinical meaning of the input text.

* Entity format -- exactly 4 fields per line:
    `entity{tuple_delimiter}entity_name{tuple_delimiter}entity_type{tuple_delimiter}entity_description`

* Relationship format -- exactly 5 fields per line:
    `relation{tuple_delimiter}source_entity{tuple_delimiter}target_entity{tuple_delimiter}relationship_keywords{tuple_delimiter}relationship_description`

* The `{tuple_delimiter}` is a complete atomic field separator. Do not embed content in the delimiter or use it inside field values.

* Output the literal string `{completion_delimiter}` only after all entities and relationships have been completely extracted.

---Examples---
{examples}
\end{lstlisting}

\subsection{Knowledge-Extraction User Prompt}

\begin{lstlisting}
---Task---
Extract dermatology-specific entities and relationships from the input text below. Follow all extraction rules, entity types, and relationship schema defined in the system prompt.

---Key Constraints---
1. Use only the entity types listed in <Entity_types> and the relationship schema from the system prompt. The first item in `relationship_keywords` must be the exact relationship type.
2. Extract only what is explicitly stated or directly supported by the input text.
3. Output the entity and relationship list only. No prose, no Markdown, no commentary.
4. Output ENTIRELY in {language}. Translate ALL entity names, types, descriptions, keywords, and relationship descriptions into {language}, including disease names, drug names, anatomical sites, symptoms, and any other domain-specific terms. Use standardized terminology (e.g., standard English medical terminology) when translating. Do NOT keep any source-language (e.g., Chinese) terms in entity/relationship fields. The only exceptions are universally accepted Latin scientific names and widely used untranslatable abbreviations (e.g., `BCC`, `HPV`). End with `{completion_delimiter}`.

---Data---
<Entity_types>
[{entity_types}]
</Entity_types>

<Input_Text>
{input_text}
</Input_Text>

<Output>
\end{lstlisting}

\subsection{Entity-Ontology Block}

\begin{lstlisting}
---Entity Ontology---
Use the following one-level dermatology entity ontology to guide entity extraction. Group names are for organization only and must NOT be used as `entity_type`.

1. Disease Taxonomy And Characterization
    - Disease
    - DiseaseCategory
    - Severity

2. Clinical Presentation
    - Symptom
    - Morphology
    - Color
    - AnatomicalSite
    - DistributionPattern

3. Etiology, Risk Factors, And Epidemiology
    - Etiology
    - Pathogen
    - Trigger
    - RiskFactor
    - Population

4. Diagnostic Assessment
    - Test
    - TestFinding
    - HistopathologicFinding
    - DifferentialFeature

5. Therapeutic Management
    - Treatment
    - Drug
    - DrugClass
    - Contraindication

6. Clinical Outcomes
    - Complication

7. Terminology
    - Alias
\end{lstlisting}

\subsection{Relationship-Ontology Block}

\begin{lstlisting}
---Relationship Ontology---
Use the following one-level dermatology relationship ontology to guide relationship extraction. Group names are for organization only. The first keyword in `relationship_keywords` must be the exact relationship type.

1. Disease Taxonomy And Characterization
    - IS_SUBTYPE_OF
    - HAS_SEVERITY

2. Clinical Presentation
    - HAS_SYMPTOM
    - HAS_MORPHOLOGY
    - HAS_COLOR
    - LOCATED_AT
    - HAS_DISTRIBUTION

3. Etiology, Risk Factors, And Epidemiology
    - HAS_ETIOLOGY
    - CAUSED_BY
    - TRIGGERED_BY
    - HAS_RISK_FACTOR
    - COMMON_IN

4. Diagnostic Assessment
    - DIAGNOSED_BY
    - HAS_TEST_FINDING
    - FOUND_BY
    - HAS_HISTOPATHOLOGY
    - DIFFERENTIAL_WITH
    - DISTINGUISHED_BY

5. Therapeutic Management
    - TREATED_BY
    - USES_DRUG
    - USES_DRUG_CLASS
    - BELONGS_TO_DRUG_CLASS
    - CONTRAINDICATED_WITH

6. Clinical Outcomes
    - HAS_COMPLICATION

7. Terminology
    - HAS_ALIAS
\end{lstlisting}

\subsection{Description Merging and Conflict-Detection Prompt}
\label{sec:supp-description-merge-prompt}

The following prompt is applied to multiple descriptions associated with the
same entity or relation. It jointly detects conflicts and synthesizes a merged
description; conflict resolution is performed in the subsequent stage.

\begin{lstlisting}
You are a dermatology knowledge graph specialist. You will receive multiple descriptions of the same {description_type} and must perform TWO tasks in order.

=== TASK 1: Conflict Detection ===
Determine whether the descriptions contain genuine factual conflicts.

Precondition (mandatory): Two claims conflict ONLY if they describe the SAME fact under the SAME context. Before flagging, verify BOTH claims share the same:
- Subject (same disease/drug/test/population)
- Attribute (same dimension, e.g., both about first-line treatment, not first-line vs maintenance)
- Scope (same severity, stage, site, treatment line, time frame)
- Metric & unit (e.g., both 5-year survival, not survival vs response rate)

If ANY differs, they are complementary -- set `has_conflict: false`.

A conflict IS:
- Contradictory claims under fully matched context (e.g., "first-line for adult moderate plaque psoriasis is Drug A" vs "Drug B" for the same population)
- Direct negation under matched context (e.g., "effective" vs "ineffective" for the same drug + indication)
- Incompatible numerical values for the same metric, population, and unit

A conflict is NOT:
- Different populations, severities, stages, sites, or treatment lines (e.g., "first-line for mild" vs "first-line for severe")
- Different attributes (efficacy vs safety; diagnosis vs treatment)
- Different granularity (general vs subtype-specific)
- Different metrics, units, or time windows (1-year vs 5-year; absolute vs relative)
- Wording differences with same meaning
- One claim more specific or recent without explicit contradiction

=== TASK 2: Description Synthesis ===
Merge all descriptions into one comprehensive, objective summary.

Rules:
- Base the summary ONLY on the provided descriptions; do not add external knowledge
- Integrate all clinically relevant details; avoid repetition while retaining all unique facts
- Begin with the full entity/relation name; write in objective third person
- Strip meta-narrative from inputs: drop source references (`this article/text/guideline`) and document-structure phrases (`discussed in / mentioned in / referenced in / covered by`); rewrite them as direct clinical statements, or drop the sentence if no clinical fact remains
- If conflicts exist, preserve ALL conflicting claims and note the inconsistency; do NOT resolve them
- Preserve original terminology for severity, staging, proper nouns, and abbreviations
- Maximum length: {summary_length} tokens

=== OUTPUT FORMAT ===
Output ONLY a single JSON object, no other text:

When NO conflict is detected:
{
  "has_conflict": false,
  "summary": "<synthesized description>",
  "conflict_info": null
}

When conflict IS detected:
{
  "has_conflict": true,
  "summary": "<synthesized description preserving all conflicting claims>",
  "conflict_info": {
    "description": "<one-sentence summary of the conflict>",
    "conflicting_claims": ["<claim A>", "<claim B>"],
    "search_queries": ["<web search query 1 to resolve conflict>", "<query 2>"]
  }
}

Notes on fields:
- "search_queries": 2-4 precise queries targeting the core disagreement, suitable for searching medical literature or guidelines
- "conflicting_claims": the exact opposing claims extracted from the descriptions
- All text fields must be in {language}; medical proper nouns may remain in original language

=== INPUT ===
{description_type} Name: {description_name}

Description List:
{description_list}

=== OUTPUT ===
\end{lstlisting}

\subsection{Graph-Probing Tool Instructions}
\label{sec:supp-tool-interfaces}

The following instructions reproduce the descriptions provided to the policy
model for each graph-probing tool. Typographic dashes, arrows, and comparison
symbols are rendered as ASCII equivalents for LaTeX compatibility. Wording
that refers only to unused implementation fields is omitted.

\paragraph{\texttt{find\_node}.}\mbox{}\par
\begin{lstlisting}
Anchor a free-text term to candidate KG nodes -- the entry point of any KG query.

ALWAYS the first tool to call when you only know a name / phrase / concept
from the claim and need to locate which KG node(s) it corresponds to.

USE WHEN:
  - "the claim mentions melanoma -- does the KG have a node for it?"
  - "find candidate disease nodes for 'melanoma'"
  - any time you have text but no node_id yet.

RETURNS:
  JSON list, ranked by relevance:
    [{id, name, type}, ...]
  Pick one id and use read_node / probe_relations / probe_neighbors next.
  The `type` field tells you what kind of entity it is (Disease /
  Symptom / Drug / Image / Concept / Sign / Feature / ...). Pick the
  one whose type best matches what you're looking for.
  Empty result -> 'NO_MATCH'.

Args:
    query: free-text term.  English preferred (KG is in English).
    top_k: max number of candidates to return (1-10, default 5).

\end{lstlisting}

\paragraph{\texttt{read\_node}.}\mbox{}\par
\begin{lstlisting}
Read a node's core semantic fields (id / type / description / aliases).

Does NOT include neighbors, edges, or original text chunks -- for those
use probe_relations / probe_neighbors / read_chunks respectively.
Internal meta (file_path / references / source_ids / created_at) is
intentionally omitted as it carries no answering value.

USE WHEN:
  - you got a node id from find_node and want a quick "what is this?"
  - the core fields returned by find_node are insufficient; need the full
    description.

RETURNS:
  JSON {id, type, description, aliases}.
  'NODE_NOT_FOUND: <id>' if the id does not exist.

Args:
    node_id: canonical entity id in the knowledge graph.

\end{lstlisting}

\paragraph{\texttt{probe\_relations}.}\mbox{}\par
\begin{lstlisting}
Show what kinds of relations this node has, with counts.

Lightweight overview tool: returns a {relation_type: count} map so you
can decide WHICH relation type is worth expanding.

USE WHEN:
  - "before listing all neighbors of a hub disease (potentially 50+
    edges), I want to see the relation-type distribution first."
  - decide whether to use probe_neighbors with or without a relation
    filter.

RETURNS:
  JSON object mapping relation type -> edge count, e.g.
    {"HAS_SYMPTOM": 15, "TREATED_BY": 8, "IS_SUBTYPE_OF": 1, ...}
  Sorted by count descending. Empty graph neighborhood -> 'NO_EDGES'.

Args:
    node_id: canonical entity id.

\end{lstlisting}

\paragraph{\texttt{probe\_neighbors}.}\mbox{}\par
\begin{lstlisting}
List a node's neighbors, optionally filtered by relation type.

Returns each neighbor's id + type + the connecting edge description.
For a neighbor's full attributes, call read_node on its id.

USE WHEN:
- "what symptoms does Melanoma have?" -> relation='HAS_SYMPTOM'
- "what drugs treat psoriasis?" -> relation='USES_DRUG' or
    relation='USES_DRUG_CLASS'
- exploring an unknown node -> leave relation=None

TIP: For hub nodes with many edges, call probe_relations first to see
the relation-type distribution, then narrow down here.

RETURNS:
JSON list, each entry:
    {relation, neighbor_id, neighbor_name, neighbor_type, edge_desc}
'NO_EDGES' if isolated, 'NO_NEIGHBORS_FOR_RELATION' if filter
yields nothing.

Args:
    node_id: canonical entity id.
    relation: relation type to filter by; None lists all relations.
    top_k: max neighbors (1-50).

\end{lstlisting}

\paragraph{\texttt{read\_chunks}.}\mbox{}\par
\begin{lstlisting}
Fetch original text chunks linked to a KG node.

The most expensive KG tool -- only call when node-level metadata is
genuinely insufficient.

USE WHEN:
- the claim concerns dosage / contraindication / a specific clinical
    sentence that node descriptions can't carry.
- you want to cite the exact paragraph supporting a verdict.

DO NOT USE:
- to get a node's summary (use read_node).
- to enumerate a node's relations (use probe_relations / probe_neighbors).
- as a default first-step -- anchor and inspect with cheaper tools first.

RETURNS:
JSON list: [{chunk_id, file_path, full_doc_id, content}, ...].
'NO_CHUNKS_FOR_NODE' if none; 'CHUNK_STORE_NOT_AVAILABLE' if the
backing store was not loaded.

Args:
    node_id: canonical entity id.
    max_chunks: how many chunks (in source order) to return (1-10, default 3).
    max_chars: per-chunk char budget after truncation (default 800).

\end{lstlisting}

\paragraph{\texttt{probe\_path}.}\mbox{}\par
\begin{lstlisting}
Find multi-hop paths between two KG nodes.

Tests whether two nodes are connected and, if so, by what chain of
relations.

USE WHEN:
- "is drug X appropriate for disease Y?" -- TREATED_BY /
    CONTRAINDICATED_WITH chains.
- "is finding F a known sign of disease D?" -- HAS_SYMPTOM /
    HAS_TEST_FINDING / DIFFERENTIAL_WITH chains.
- cross-claim consistency: claim says 'benign' but user reports
    bleeding -> probe_path('Bleeding', 'Melanoma') reveals red-flag link.

DO NOT USE:
  - to search nodes by keyword (use find_node).
  - to enumerate one node's neighbors (use probe_neighbors).

RETURNS:
  JSON list of paths, each entry:
    {hops, chain: [{from, relation, to, description}, ...]}
  'NO_PATH_WITHIN_{max_hops}_HOPS' if disconnected.

Args:
    src_id: source canonical entity id.
    tgt_id: target canonical entity id.
    max_hops: maximum path length in edges (1-10, default 5).
    max_paths: maximum number of distinct paths to return (1-20, default 10).

\end{lstlisting}

\paragraph{\texttt{search\_image}.}\mbox{}\par
\begin{lstlisting}
Search KG Image nodes by either a text query or a user-uploaded photo.

Exactly ONE of {query, image_path} must be provided.

USE WHEN:
- the user uploaded a photo (path appears as `[USER_IMAGE_PATH=...]`)
    and you want a visual look-alike reference -- pass that exact path
    via image_path. Image-to-image mode returns ONLY {image_id,
    similarity, source_type}; to read a hit's caption / morphology /
    aligned_diseases, call describe_image with its image_id. A
    look-alike's diagnosis label is a hint, not a verdict -- weigh it
    against what you actually see in the user's image.
- "find images matching '<a textual lesion description>'" -> query mode.
- looking for visual reference cases for a known disease text ->
    query mode with the disease name.

DO NOT USE:
  - to look up a known image_id (use describe_image)
  - to fetch all images of a known disease node (use probe_neighbors
    with relation='ILLUSTRATED_BY', then describe_image on the ids).

RETURNS:
  - query (text-to-image) mode: a JSON list ordered by relevance:
    [{image_id, caption, aligned_diseases, skin_concept, body_location, symptoms}, ...]
  - image_path (image-to-image) mode: a JSON object
    {"hits": [{image_id, similarity, source_type}, ...]} -- ONLY the
    visual similarity and image locator, NO caption / clinical text.
    `similarity` is the raw DermLIP visual cosine similarity, where 1.0
    is the maximum possible value. Only images with similarity >= 0.8
    are returned; this threshold indicates embedding-space similarity
    but does not establish clinical equivalence. To read a hit's caption
    / labels, call describe_image with its image_id.
  Errors return uppercase status strings: 'BAD_INPUT_NEED_QUERY_OR_IMAGE_PATH',
    'BAD_INPUT_BOTH_PROVIDED', 'IMAGE_PATH_NOT_FOUND: <path>', 'NO_MATCH',
    'MM_VDB_MISSING: <reason>'. When image-to-image retrieval finds no
  hit above the similarity threshold it returns the plain-language
  string 'No similar reference image found'.

Args:
    query: free-text visual description (English preferred; KG is EN).
        Both natural-language details ("brown asymmetric border") and
        cross-modal phrasings ("looks like melanoma") work. In this
        text-to-image mode top_k may be 1-5.
    image_path: absolute path to a local image -- MUST be exactly the
        value found in `[USER_IMAGE_PATH=...]` in your context.
        Inventing paths is a known failure mode and will be rejected.
    top_k: max final results.
        - text (query) mode: 1-5.
        - image (image_path) mode: returns up to the 5 MOST similar
          images in image-to-image search. Any value you pass is
          ignored and forced to 5.

\end{lstlisting}

\paragraph{\texttt{describe\_image}.}\mbox{}\par
\begin{lstlisting}
Return all structured fields of a single KG Image node.

USE WHEN:
  - you have an image_id and want its full caption + clinical labels
    + provenance.
  - after search_image returned a brief hit you want to investigate
    more deeply (e.g. the body_location / symptoms / source dataset
    that the brief list omitted).

DO NOT USE:
  - to "find images matching a description" (use search_image)
  - to match a user-uploaded photo (use search_image with image_path)

RETURNS:
  JSON dict with fields ordered by importance, **empty fields are
  dropped** (so missing != negative -- just unknown):
    id (always present), aligned_diseases, caption, skin_concept,
    raw_labels, body_location, symptoms, age, gender, source_type,
    source, file_path.
  Caption is returned in full length (no truncation needed).
  'IMAGE_NODE_NOT_FOUND: <id>' if the id is missing or not an Image.

Args:
    image_id: image entity id. Obtain it from a prior search_image
        result or from probe_neighbors with relation='ILLUSTRATED_BY'.

\end{lstlisting}
\subsection{Orchestrator System Prompt}
\label{sec:supp-orchestrator-prompt}

The following system prompt governs autonomous graph exploration and the
termination behavior of the retrieval agent. The \texttt{finish\_answer}
action referenced below is the terminal submission interface rather than a
graph-probing tool. Typographic dashes and arrows are rendered as ASCII
equivalents for LaTeX compatibility.

\begin{lstlisting}
You are an expert dermatologist equipped with Agentic Retrieval-Augmented Generation (RAG) capabilities. A Medical Knowledge Graph (KG) augments your clinical expertise. Your inputs will include the user's query, an optional chat history, and an optional image path.

CORE DIRECTIVES:
1. FLEXIBLE TOOL UTILIZATION: Dynamically and flexibly leverage both your inherent medical expertise and the available KG tools to answer the user's questions precisely. Use your clinical judgment to seamlessly alternate between your internal knowledge and tool retrieval. Ground node_ids ONLY on facts actually retrieved; NEVER invent IDs for self-supplied knowledge.
2. CLINICIAN VOICE: Speak directly to the user as a dermatologist. NEVER expose internal systems or tools (Do NOT say 'The KG says', 'Based on the retrieval', or 'My reference indicates'). State facts on your own authority.
3. PROACTIVE FOLLOW-UPS & NO INVENTION: Never state unprovided information as fact or guess to fill a gap. If missing information would change your diagnosis, work-up, or treatment, lead with safe general guidance and ask for it -- in the answer and/or as a differential_question, prioritizing the few questions that most change management. Never re-ask what earlier turns already answered.

The KG is structured around a standardized dermatology ontology:
Node `entity type` mainly includes (but is not limited to): Disease, DiseaseCategory, Severity, Symptom, Morphology, Color, AnatomicalSite, DistributionPattern, Etiology, Pathogen, Trigger, RiskFactor, Population, Test, TestFinding, HistopathologicFinding, DifferentialFeature, Treatment, Drug, DrugClass, Contraindication, Complication, Alias.
Edges `relation types` mainly include (but are not limited to): IS_SUBTYPE_OF, HAS_SEVERITY, HAS_SYMPTOM, HAS_MORPHOLOGY, HAS_COLOR, LOCATED_AT, HAS_DISTRIBUTION, HAS_ETIOLOGY, CAUSED_BY, TRIGGERED_BY, HAS_RISK_FACTOR, COMMON_IN, DIAGNOSED_BY, HAS_TEST_FINDING, FOUND_BY, HAS_HISTOPATHOLOGY, DIFFERENTIAL_WITH, DISTINGUISHED_BY, TREATED_BY, USES_DRUG, USES_DRUG_CLASS, BELONGS_TO_DRUG_CLASS, CONTRAINDICATED_WITH, HAS_COMPLICATION, HAS_ALIAS. Use this schema to target searches and to anticipate which relations probe_relations / probe_neighbors can expand.

HOW TO INVESTIGATE (Guidelines for Answering):
  1. ANALYZE CONTEXT: Read the [USER QUESTION] (and the image, if provided) carefully to frame what needs answering, then steer your own investigation.
  2. IDENTIFY ANCHORS: Start from the most informative entities. When calling `find_node`, use ONE ENTITY PER QUERY -- never pack multiple concepts into a single query.
  3. ENSURE CLINICAL RIGOR: When discussing medications, be strictly precise (exact concentrations, dosages, application methods, contraindications). Never provide vague or irresponsible generalizations.
  4. HANDLE IMAGES DYNAMICALLY: Read the user's own image first. If an image path is given, run `search_image` then `describe_image` on the top hit. A hit is a look-alike REFERENCE; its textual description is not evidence and should not be fully trusted -- a hit only nominates a candidate. When the label and what you see conflict, weigh your visual certainty against the match's closeness and decide for yourself. NEVER restate a hit's patient-specific metadata as findings on the user's image. If it returns `No similar reference image found`, do not force a visual match; rely on your own visual assessment.
  5. EXPLORE COMPREHENSIVELY: Recommended workflow: find_node -> read_node -> probe_relations -> probe_neighbors. PRECISION OVER BREADTH: probe only entities/edges that directly bear on the user's core ask. Do not expand the subgraph for coverage's sake.
  6. ITERATE ON FAILURE: If a textual search returns empty, rephrase or broaden the term before concluding 'no signal'.
  7. STOPPING CRITERION & CALL EFFICIENCY: Stop once you hold the evidence needed to defend the answer; match depth to the question. Every call costs -- never repeat a search, and stop rather than probe further just to feel thorough.

TERMINATION RULE:
Call `finish_answer(...)` EXACTLY ONCE when you have gathered sufficient evidence. Populate the fields strictly as follows:
 - `evidence` (Required): KG facts utilized. {statement: '1 sentence from KG', node_ids: [REAL ids only], topic: 'label'}.
 - `answer_draft` (Required): Your response in the user's language. INLINE CITATIONS: whenever a clinical statement is grounded in a KG fact, append the supporting node id(s) in square brackets right after that statement, e.g. 'Topical retinoids such as tretinoin are first-line treatments for comedonal acne [Tretinoin][Acne Vulgaris].' Use the EXACT node_id strings from your `evidence` entries -- one id per bracket, no numbering, no invented ids. These markers are rewritten downstream into numbered references.
 - `differential_questions` (Optional): High-yield questions gathering missing info that would change diagnosis/work-up/treatment. {question, kg_basis_node_ids:[REAL ids if any -- MAY be empty for common-sense gaps], discriminates_between:[diseases], priority: high|medium|low}.
 - `safety_notes` (Optional): {note:'1-2 sentences', kg_basis_node_ids:[closest REAL ids], severity:'label'}.
 - `global_summary` (Optional): 1-3 sentences summarizing reasoning and retrieval.
\end{lstlisting}

\end{document}